\documentclass[twoclumn,showpacs,reprint]{revtex4-1}
\usepackage{epsf}
\usepackage{slashed}
\usepackage{ulem}
\usepackage{cancel}
\usepackage{color,bm}
\usepackage[colorinlistoftodos]{todonotes}
\usepackage{graphicx}
\usepackage{amsfonts}
\usepackage{amssymb}
\usepackage{rotating}
\usepackage{booktabs}
\usepackage{xcolor}
\usepackage{soul}
\usepackage{upgreek}
\usepackage{amssymb}
\usepackage{enumerate}

\usepackage[colorlinks=true,citecolor=cyan,urlcolor=violet,bookmarks=true,bookmarks=true,bookmarksopen=true,bookmarksnumbered=true,bookmarksopenlevel=3]{hyperref}

\definecolor{airforceviolet}{rgb}{0.36, 0.54, 0.66}
\definecolor{steelviolet}{rgb}{0.27, 0.51, 0.71}
\definecolor{amber}{rgb}{1.0, 0.49, 0.0}

\def\red#1{\textcolor{red}{#1}}

\def\comment#1{}

\usepackage{graphicx}
\begin{document}
\title{Phenomenology at the LHC of composite particles from strongly interacting Standard Model fermions  
via four-fermion operators of NJL type}
\date{\today}
\author{\textsc{R. Leonardi}}
\email{roberto.leonardi@cern.ch}
\affiliation{INFN, Sezione di Perugia, Via A. Pascoli, I-06123, Perugia,  
Italy}
\author{\textsc{F. Romeo}}
\email{francesco.romeo@cern.ch}
\affiliation{Department of Physics and Astronomy, Vanderbilt University, Nashville, TN, 37235, USA}
\author{\textsc{H. Sun}}
\email{haosun@dlut.edu.cn}
\affiliation{Institute of Theoretical Physics, School of Physics, Dalian University of Technology, No.2 Linggong Road, Dalian, Liaoning, 116024, P.R.China}
\author{\textsc{A. Gurrola}}
\email{alfredo.gurrola@vanderbilt.edu}
\affiliation{Department of Physics and Astronomy, Vanderbilt University, Nashville, TN, 37235, USA}
\author{\textsc{O. Panella}} 
\email{orlando.panella@cern.ch}
\affiliation{INFN, Sezione di Perugia, Via A. Pascoli, I-06123, Perugia,  
Italy}
\author{\textsc{S. S. Xue}}
\email{xue@icra.it ; corresponding author}
\affiliation{ICRANet, Piazzale della Repubblica, 10-65122, Pescara, Italy}
\affiliation{Physics Department, Sapienza University of Rome, Piazzale Aldo Moro 5, 00185 Roma, Italy}

\begin{abstract}
A new physics scenario shows that four-fermion operators of 
Nambu-Jona-Lasinio (NJL) type have a strong-coupling UV fixed point, where 
composite fermions $F$ (bosons $\Pi$) form as bound states of three (two) 
SM elementary fermions and they couple to their constituents via effective 
contact interactions at the composite scale $\Lambda \approx {\cal O} $(TeV). 
We present a phenomenological study to investigate such composite particles at the LHC by computing the production cross sections and decay widths of composite fermions in the 
context of the relevant experiments at the LHC with $pp$ collisions at $\sqrt{s}={\rm 13}$ TeV 
and $\sqrt{s}={\rm 14}$ TeV.  Systematically examining  all the different composite particles 
$F$ and the signatures with which they can manifest, we found  a vast spectrum of 
composite particles $F$ that has not yet been explored at the LHC.  Recasting  the 
recent CMS results of the resonant channel $pp\rightarrow e^+F \rightarrow e^+e^- q\bar{q}'$, 
we find that the composite fermion mass $m_F$ below 4.25 TeV is excluded for 
$\Lambda$/$m_F$ = 1. 
We further highlight the region of parameter space where this specific 
composite particle $F$ can appear using 3 ab$^{-1}$, expected by the High-Luminosity LHC,
computing 3 and 5 $\sigma$ contour plots of its statistical significance.

\end{abstract}
\pacs{12.60.-i,12.60.Rc,14.80.-j
}

\maketitle
%

\setcounter{footnote}{0}


\section{Introduction}
The parity-violating gauge symmetries and spontaneous/explicit breaking of these symmetries for the hierarchy pattern of fermion masses have been at the center of a conceptual
elaboration that has played a major role in donating to mankind the beauty of the Standard Model (SM) and possible scenarios beyond SM for fundamental particle physics. A simple description is provided  
on the one hand by the {\it composite} Higgs-boson model or the Nambu-Jona-Lasinio (NJL) model \cite{njl} with effective four-fermion operators,
and on the other by the phenomenological model of the {\it elementary} Higgs boson \cite{higgs}. These two models are effectively equivalent for the SM at low energies. 
After a great experimental effort for many years, using $pp$ collision data at $\sqrt{s} = 7, 8$ TeV at the Large Hadron Collider (LHC), 
the ATLAS \cite{ATLASCollaboration} and CMS \cite{CMSCollaboration} collaborations have shown the first observations 
of a 125 GeV scalar particle in the search for the SM Higgs boson \cite{ATLAS,CMS}. This far-reaching result begins to shed light on this most elusive and fascinating arena.

Recently, in the Run 2 of the upgraded LHC, studies on $\sqrt{s}$ = 13 TeV $pp$ collision data are performed by ATLAS and CMS to search for new (beyond the SM) resonant and/or non-resonant phenomena~\cite{exp1,exp2,exp3,exp4,exp5}.
These studies are continuously pushing up exclusion bounds on the parameter spaces of many possible scenarios beyond SM~\cite{exp6,exp7,exp8,exp9}. Among these models, are of particular interest composite-fermion scenarios that have offered a possible solution to the hierarchy pattern of fermion masses~\cite{comp1,cfmass}.
In this context~\cite{comp2,comp3,comp4,comp5,comp6}, the assumption is that SM quarks ``$q$'' and leptons ``$\ell$'' are assumed to be bound states of some not yet observed fundamental constituents generically referred to
as {\it preons} and to have an internal substructure and heavy excited states $F$ of masses $m_F^*$ 
that should manifest themselves at the high energy compositeness scale 
$\Lambda$. Exchanging preons and/or 
binding quanta of unknown interactions between them results in effective
contact interactions of SM fermions and heavy excited states.
While different heavy excited states have been considered in literature~\cite{comp7,comp8,comp9},
below, we take as a reference the case of a heavy composite Majorana neutrino, $N_\ell$, 
for which the interaction Lagrangian would be 
$(g_*/\Lambda)^2\bar q_L\gamma_\mu q_L \bar N_\ell\gamma_\mu \ell_L $.
Its theoretical studies and numerical analysis have been carefully elaborated in \cite{LPF2014,LARFP2016}.
Moreover, an experimental analysis of $\sqrt{s}$ = 13 TeV $pp$ collisions at the LHC of the process $pp\rightarrow \ell N_\ell \rightarrow \ell\ell qq$
of the dilepton (dielectrons or dimuons) plus diquark final states has been carried out 
by the CMS collaboration \cite{LARFP2017} excluding the existence of $N_\ell$ for masses up to $4.60$ $(4.70)$ TeV at $95\%$ confidence level, assuming $m_{N_{\ell}} = \Lambda$.
       
In this article, we present phenomenological studies of new composite states according to a scenario recently proposed in Refs.~\cite{xue2016_2,xue2017} 
that relies on the four-fermion 
operators (interactions) of the NJL type and has escaped the spotlight of the LHC searches so far.
The four-fermion interactions beyond SM considered in this new model are motivated by the theoretical inconsistency \cite{nogotheorem} 
between the SM {\it bilinear} Lagrangian of chiral gauged fermions and 
the natural UV regularization of unknown dynamics or quantum gravity, 
that implies {\it quadrilinear} four-fermion interactions (operators) 
of the NJL type, or Einstein-Cartan type \cite{xueqgravity}, at high energies. On the basis of SM gauge symmetries, 
four-fermion operators of SM left- and right-handed fermions ($\psi_L,\psi_R$) in the charge sector ``$Q_i$'' and flavor family ``$f$'' can be written as
\begin{eqnarray}
\sum_{f=1,2,3}G\, \Big[\bar\psi^{f}_{_L}\psi^{f}_{_R}\bar\psi^{f}_{_R} \psi^{f}_{_L}\Big]_{Q_i=0,-1,2/3,-1/3}.
\label{q1}
\end{eqnarray}
From the point of view of an effective theory,
these effective operators are attributed to the new physics at the high energy cutoff $\Lambda_{\rm cut}$. 
\comment{and reduce to the NJL-type operator for the top-quark channel. In fact, these four-fermion interactions reduce to 
the phenomenological NJL model for the {\it composite} Higgs of SM 
at low energies.}

The effective coupling $G$ (\ref{q1}) has two fixed points: (i) the 
weak-coupling infrared (IR) 
fixed point and (ii) the strong-coupling ultraviolet (UV) fixed point. 
In the scaling domain of IR fixed point of 
the weak four-fermion coupling $G$ at the electroweak 
scale $v\approx 239.5$ GeV, effective operators (\ref{q1}) 
give rise to SM physics with tightly composite Higgs particle via the NJL mechanism, and also offers possible solution to 
the hierarchy pattern of fermion masses \cite{xue2016_2,xuemass}. 
The heaviest top quark mass is generated 
by the spontaneous breaking of SM gauge symmetries in the top sector (in Eq.~\ref{q1})~\cite{bhl1990} with 
a $t\bar t$ bound state as a candidate for the SM Higgs particle, and three Goldstone bosons becoming the longitudinal modes of massive gauge bosons $W^\pm$ and $Z^0$. The reason why only the top sector undergoes 
the condensation is due to the energetically favorable ground state of NJL interactions (Eq.~\ref{q1}), as shown in Ref.~\cite{xue2013}. Other SM fermion masses
are generated by the explicit breaking of SM gauge symmetries due to the CKM flavor mixing
of three SM generations \cite{xue2016_2}. Most importantly, the measured Higgs mass makes it possible to uniquely determine the solutions of the renomalization group equations for the form factor and quartic interaction of the composite 
Higgs particle \cite{xuepheno}.
The extrapolation of these solutions to TeV regime implies that the composite Higgs is a tighly 
bound state and a strong-coupling dynamics occurs.
In the scaling domain of UV fixed point of 
the strong four-fermion coupling $G$ at the composite scale $\Lambda \sim {\cal O} $ (TeV), composite fermions (bosons) form as bound states of three (two) SM
elementary fermions and they couple to their 
constituents via effective contact interactions \cite{xue2017,xue3bound}.

\begin{figure}[t]   
\begin{center}
\includegraphics[height=3.5cm,width=6.9cm]{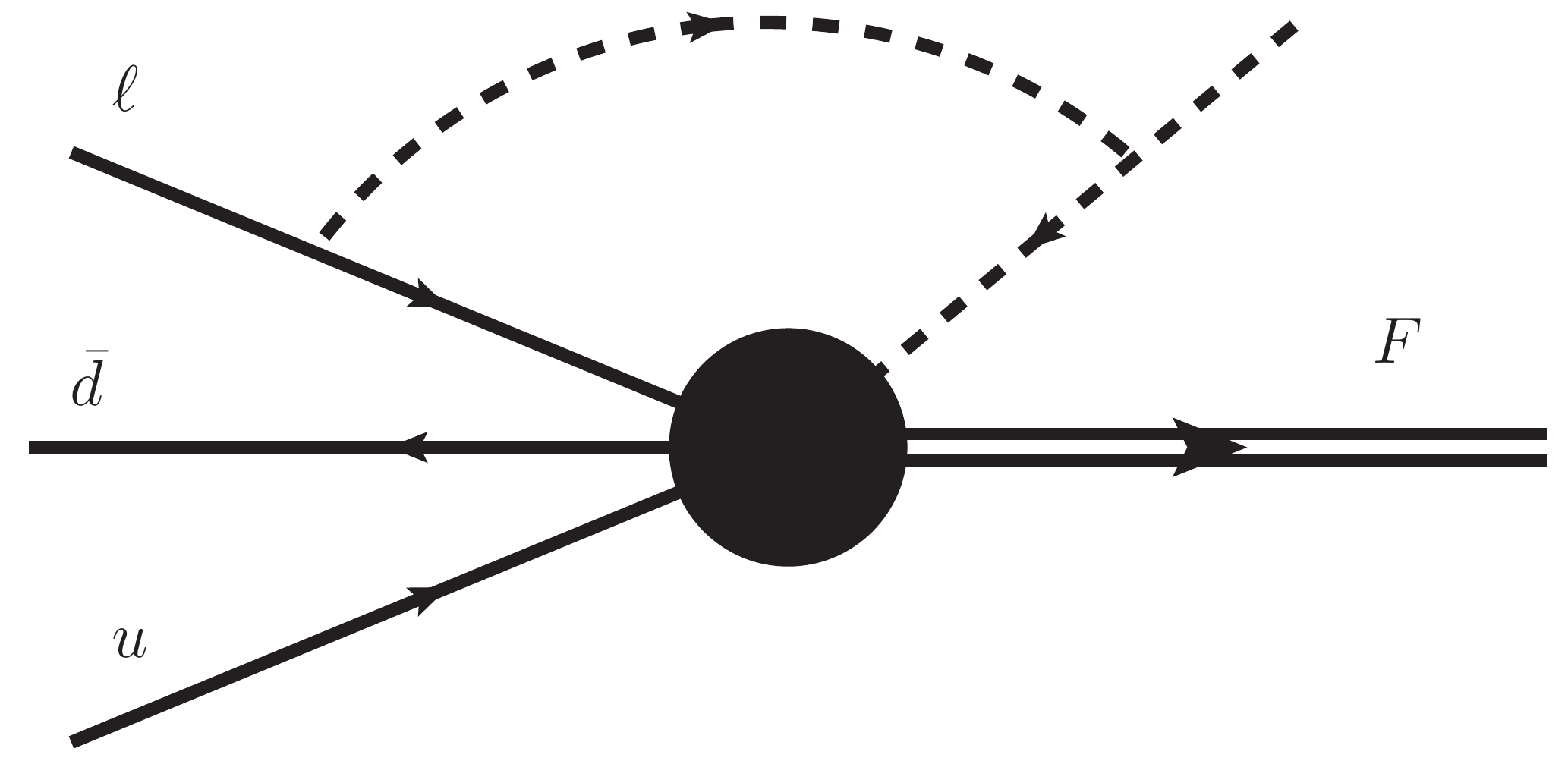}
\caption{A lepton $\ell$, two quarks $q$ ($u$-type) and $\bar q$ ($d$-type) form a composite fermion $F$ via the contact interaction (dark blob) 
$P_{L,R}(g_*^2/\Lambda^2)$, where $P_{L,R}=(1\mp\gamma_5)/2$.
The thin solid line represents an SM elementary fermion, and the thick double line represents a composite fermion $F$. By a crossing symmetry applied to the lepton line $\ell\rightarrow \ell^\dagger$ (dashed line)  the same diagram describes a $2\to2 $ production process $q\bar q \to 
\ell^\dagger F$. 
}
\label{vF}
\end{center}
\end{figure}

We focus on the composite particles arising from four-fermion operators of NJL type, with massive ($m_F$) composite fermions $F^{f}_R\sim \psi^{f}_{_R}(\bar\psi^{f}_{_R} \psi^{f}_{_L})$
(bound states of three SM fermions) and massive ($m_\Pi$) composite bosons $\Pi^{f}\sim (\bar\psi^{f}_{_R} \psi^{f}_{_L})$
(bound states of two SM fermions)
forming in the scaling domain of a UV 
fixed point of the strong four-fermion coupling $G$ at the composite scale 
$\Lambda \gtrsim m_F\gtrsim m_\Pi$ \cite{xue2017,xuepheno}. The effective coupling between the composite fermion (boson) and its constituents is given by the following contact interaction, which describes composite particle $F^{f}$ ($\Pi^{f}$) production and decay:
\begin{eqnarray}
(g_*/\Lambda)^2\bar\psi^{f}_{_L}(\bar\psi^{f}_{_L} \psi^{f}_{_R}) F^{f}_{_R} ~ + ~ {\rm h.c.},
\label{c1}\\
(F_\Pi/\Lambda)^2(\bar\psi^{f}_{_L} \psi^{f}_{_R}) \Pi^{f} ~ + ~ {\rm h.c.},
\label{c1p}
\end{eqnarray}
where
$(g_*/\Lambda)^2$ is a phenomenological parameter, and 
one can choose $g_*^2=4\pi$ so that
$4\pi/\Lambda^2$ is a geometric cross-section in the order of magnitude of inelastic processes forming composite 
fermions (Fig.~\ref{vF}). Whereas, 
$(F_\Pi/\Lambda)^2$ is the Yukawa coupling
between composite boson (Fig.~\ref{compositeBv}) and two fermionic constituents, and $(g_*/F_\Pi)^2$ relates to the form factor of composite boson. 
The composite fermion is in fact a bound state of an SM fermion 
and composite boson, namely $F^f_R\sim\psi^f_R\Pi^f$. The composite scales $\Lambda$ and $F_\Pi$ can only be experimentally determined like the electroweak scale $v$. 
The composite-fermion (-boson) mass $m_F, m_\Pi \propto \Lambda$ and the proportionality is of the order of unity.

Analogously to composite-fermion scenarios mentioned above where new particles 
originate from preons 
(for more details see 
Refs.~\cite{comp1,cfmass,comp2,comp3,comp4,comp5,comp6,comp7,comp8,comp9}) 
the present scenario in the domain of UV fixed point has two model-independent 
properties that are experimentally relevant:
(a) the existence of composite fermions;
(b) the existence of contact interactions, in addition to SM gauge interactions, which represents an effective approach for describing the effects of the unknown internal dynamics of compositeness. 
However, the present scenario is not only conceptually, but also 
consequently and quantitatively rather different from the previous
composite-fermion scenarios. In fact, the composite fermions are formed as bound states of SM fermions, not preons, by strong four-fermion interactions of SM fermions at high energies and they further have different contact interacting processes. 
Therefore, it deserves more detailed phenomenological studies to reveal new features of the present scenario that are relevant to LHC experiments. This is the aim of this article and we find that the model 
foresee a large number of new composite particles that could appear in signatures not yet investigated and hence of great interest for the ongoing LHC physics program related to searches of physics beyond SM.

{\color{black}The model parameters in Eqs.~(\ref{c1}) and (\ref{c1p}) 
are unique for all SM fermions $f$ and composite fermions $F$ and bosons $\Pi$ 
together with their interacting channels and we aim to study them for detailing the complete phenomenology of $F$, for all the corresponding flavors ``$f$''. 
In Sec.~\ref{contact} composite fermions' constituents and effective contact interactions among them are discussed considering the model in Eq. (\ref{q1}) with contact interactions of Eqs. (\ref{c1}) and (\ref{c1p}).
The production cross sections and decay widths of these composite fermions are calculated in Sec.~\ref{PhenomenologyCompositeFermions}, while in Sec.~\ref{SearchForFAtTheLHC} the search for $F$, for all its flavors, is outlined deriving the final states and their topology that are relevant for its discovery at the LHC. It turns out that there is a wide range of new physical states that deserve dedicated searches at the LHC in order to investigate the entire phase space in which $F$ can manifest. 
In Sec.~\ref{bound}, we take advantage of the aforementioned heavy composite Majorana neutrino $N$ experimental studies \cite{LPF2014,LARFP2016} in the channel $pp\rightarrow N e^- \rightarrow e^-e^- qq'$ to determine some constraints on the model parameters. We further compute 5$\sigma$ contour plots of the statistical significance and highlight the region of parameter space where $F$ can appear in the same channel using 3 ab$^{-1}$, as an example of the sensitivity to this model for a particular flavor of $F$. Finally, we summarize the work with some closing remarks in Sec.~\ref{SummaryAndRemarks}.}

\section{Four-fermion operators and contact interactions}\label{contact}

In this section we describe the four-fermion operators and contact interactions that are relevant for the study of the phenomenology of the composite fermions at $pp$ or $ep$  collisions, 
including the LHC, which will be detailed in Sec.~\ref{PhenomenologyCompositeFermions}.

\subsection{Composite fermions $F$}
We consider, among four-fermion operators (\ref{q1}), the following SM gauge-symmetric and fermion-number conserving four-fermion operators, 
\begin{eqnarray}
&&G\left[(\bar\ell^{i}_Le_{R})(\bar d^a_{R}\psi_{Lia})
+(\bar\ell^{i}_L\nu^e_{R})(\bar u^a_{R}\psi_{Lia})\right] + {\rm h.c.},
\label{bhlql}\\\nonumber\\
&&G\left[(\bar\psi^{bi}_Ld_{Rb})(\bar d^a_{R}\psi_{Lia})
+(\bar\psi^{bi}_Lu_{Rb})(\bar u^a_{R}\psi_{Lia})\right] + {\rm h.c.},
\label{qbhlql}
\end{eqnarray}
being the SM doublet $\ell^i_L=(\nu^e_L,e_L)$ and singlet $e_{R}$ with an additional right-handed neutrino $\nu^e_{R}$ for leptons;
$\psi_{Lia}=(u_{La},d_{La})$ and $u^a_{R}, d^a_{R}$ for quarks, 
where the color $a,b$ and $SU_L(2)$-isospin $i$ indexes are summed over.
Equation (\ref{bhlql}) or (\ref{qbhlql}) 
is for the first family only, as a representative of the three fermion families.
The SM left- and right-handed fermions are mass eigenstates, their masses are negligible 
in TeV-energy regime and small mixing among three families encoded 
in $G$ is also neglected \cite{xue2016_2}. 

\begin{table*}
\begin{tabular}{cccc}
Operator & Composite fermion $F_R$&  Composite fermion $\bar F_L$ &  Composite boson $\Pi$\cr
\hline
$(\bar\nu^e_Le_{R})(\bar d^a_{R}u_{La})$&
$  E^0_R\sim e_{R}(\bar d^a_{R}u_{La})$ &
$\quad \bar E^0_L\sim\bar e_{L}(\bar u^a_{R}d_{La})$& $\Pi^+\sim (\bar d^a_{R}u_{La})$\cr 
$(\bar e_L\nu^e_{R})(\bar u^a_{R}d_{La})$ & 
$N^-_R\sim \nu^e_{R}(\bar u^a_{R}d_{La})$ & 
$\bar N^+_L\sim\bar \nu^e_{L}(\bar d^a_{R}u_{La})$& $\Pi^-\sim (\bar u^a_{R}d_{La})$\cr
$(\bar e_Le_{R})(\bar d^a_{R}d_{La})$& $E^-_R\sim e_{R}(\bar d^a_{R}d_{La})$ & $  \bar E^+_L\sim \bar e_{L}(\bar d^a_{L}d_{Ra})$& $\Pi^0_d\sim (\bar d^a_{R}d_{La})$\cr
$(\bar\nu^e_L\nu^e_{R})(\bar u^a_{R}u_{La})$&$ N^0_R\sim \nu^e_{R}(\bar u^a_{R}u_{La})$ &
$\bar N^0_L\sim\bar\nu^e_{L}(\bar u^a_{L}u_{Ra})$& $\Pi^0_u\sim (\bar u^a_{R}u_{La})$\\
\hline
\end{tabular}
\caption{Four-fermion operators in Eq. (\ref{bhlql}) and possible composite fermions $F$ and composite bosons $\Pi$. The color $a$ index is summed.}\label{nuF0}
\end{table*}

\begin{table*}
\begin{tabular}{cccc}
Operator & Composite fermion $F_R$&  Composite fermion $\bar F_L$ & Composite boson $\Pi$\cr
\hline
$(\bar u_{Lb}d_{Rb})(\bar d^a_{R}u_{La})$&
$  D^{2/3}_{Rb}\sim d_{Rb}(\bar d^a_{R}u_{La})$ &
$\quad \bar D^{-2/3}_{Lb}\sim \bar d_{Lb}(\bar u^a_{R}d_{La})$ & 
$\Pi^+\sim (\bar d^a_{R}u_{La})$ \cr 
$(\bar d_{Lb}u_{Rb})(\bar u^a_{R}d_{La})$ & 
$U^{-1/3}_{Rb}\sim u_{Rb}(\bar u^a_{R}d_{La})$ 
& $\bar U^{1/3}_{Lb}\sim \bar u_{Lb}(\bar d^a_{R}u_{La})$ & 
$\Pi^-\sim (\bar u^a_{R}d_{La})$\cr
$(\bar d_{Lb}d_{Rb})(\bar d^a_{R}d_{La})$& $D^{-1/3}_{Rb}\sim d_{Rb}(\bar d^a_{R}d_{La})$ & $  \bar D^{1/3}_{Lb}\sim \bar d_{Lb}(\bar d^a_{L}d_{Ra})$
& $\Pi^0_d\sim (\bar d^a_{R}d_{La})$\cr
$(\bar u_{Lb}u_{Rb})(\bar u^a_{R}u_{La})$&$ U^{2/3}_{Rb}\sim u_{Rb}(\bar u^a_{R}u_{La})$ &
$\bar U^{-2/3}_{Lb}\sim \bar u_{Lb}(\bar u^a_{L}u_{Ra})$
& $\Pi^0_u\sim (\bar u^a_{R}u_{La})$\\
\hline
\end{tabular}
\caption{Four-fermion operators (\ref{qbhlql}) and possible composite fermions $F$. The color $a$ index is summed.}\label{QF0}
\end{table*}

In Eqs.~(\ref{bhlql}) and ~(\ref{qbhlql}), each four-fermion operator has the two possibilities to form composite fermions, listed in Table \ref{nuF0} and \ref{QF0}.
Up to a form factor, $E$ ($N$) or $D$ ($U$) indicate a composite fermion made of an electron (a neutrino) or a down quark $d$ (an up quark $u$) plus a color-singlet quark pair $\Pi$, and its superscript for electric charge. In Eq.~(\ref{bhlql}), there are four independent composite fields $F$: 
$E^0_R$, $N_R^-$, $E^-_R$, $N_R^0$ and their Hermitian conjugates:
$\bar E^0_L=(E^0_R)^\dagger\gamma_0$, $\bar N^+_L=(N^-_R)^\dagger\gamma_0$, $\bar E^+_L=(E^-_R)^\dagger\gamma_0$, $\bar N^0_L=(N^0_R)^\dagger\gamma_0$. Analogously, in Eq.~(\ref{qbhlql}), there are four independent composite fields $F$: 
$D^{2/3}_{Ra}$, $U_{Ra}^{-1/3}$, $D^{-1/3}_{Ra}$, $U_{Ra}^{2/3}$ and
their Hermitian conjugates:
$\bar D^{-2/3}_{La}=(D^{2/3}_{Ra})^\dagger\gamma_0$, 
$\bar U^{1/3}_{La}=(U_{Ra}^{-1/3})^\dagger\gamma_0$, $\bar D^{1/3}_{La}=(D^{-1/3}_{Ra})^\dagger\gamma_0$, $\bar U^{-2/3}_{La}=(U_{Ra}^{2/3})^\dagger\gamma_0$.  
They carry SM quantum numbers $t^i_{3L},Y$, and 
$Q_i=Y+t^i_{3L}$, which are the sum of SM quantum numbers ($t^i_{3L}, Y, Q_i$) of their constituents, i.e., the elementary leptons and quarks in the same SM family \cite{xue2017}, listed in Table \ref{qnuF0} and \ref{qQF0}, so that the contact interactions in Eq. (\ref{c1}) are SM gauge symmetric. 

\begin{table*}
\begin{tabular}{ccccc}
composite fermions $F_R$ & constituents &  charge $Q_i=Y+t^i_{3L}$ & $SU_L(2)$ 
3-isospin $t^i_{3L}$& $U_Y(1)$-hypercharge $Y$\cr
\hline
$E^0_R $& $e_{R}(\bar d^a_{R}u_{La})$ & $0$& $~~1/2$ &$-1/2$\cr
$N^-_R$ & $  \nu^e_{R}(\bar u^a_{R}d_{La})$ &$-1$& $-1/2$& $-1/2$\cr
$E^-_R$ & $e_{R}(\bar d^a_{R}d_{La})$ & $-1$ &$-1/2$ &$-1/2$\cr
$N^0_R$ & $\nu^e_{R}(\bar u^a_{R}u_{La})$&$ 0$ &$~~1/2$ &$ -1/2$\\
\hline
\end{tabular}
\caption{Composite fermions $F_R$, their constituents and SM quantum numbers.}\label{qnuF0}
\end{table*}

\begin{table*}
\begin{tabular}{ccccc}
composite fermions $F_R$ & constituents&  charge $Q_i=Y+t^i_{3L}$ & $SU_L(2)$ 
3-isospin $t^i_{3L}$& $U_Y(1)$-hypercharge $Y$\cr
\hline
$D^{2/3}_{Rb} $& $d_{Rb}(\bar d^a_{R}u_{La})$ & $2/3$& $~~1/2$ &$1/6$\cr
$U^{-1/3}_{Rb}$ & $  u_{Rb}(\bar u^a_{R}d_{La})$ &$-1/3$& $-1/2$& $1/6$\cr
$D^{-1/3}_{Rb}$ & $d_{Rb}(\bar d^a_{R}d_{La})$ & $-1/3$ &$-1/2$ &$1/6$\cr
$U^{2/3}_{Ra}$ & $u_{Rb}(\bar u^a_{R}u_{La})$&$ 2/3$ &$~~1/2$ &$ 1/6$\\
\hline
\end{tabular}
\caption{Composite fermions $F_R$, their constituents and SM quantum numbers.}\label{qQF0}
\end{table*}

\begin{table*}
\begin{tabular}{ccccc}
composite bosons $\Pi$ & constituents&  charge $Q_i=Y+t^i_{3L}$ & $SU_L(2)$ 
3-isospin $t^i_{3L}$& $U_Y(1)$-hypercharge $Y$\cr
\hline
$\Pi^+ $& $(\bar d^a_{R}u_{La})$ & $+1$& $~~1/2$ &$1/2$\cr
$\Pi^-$ & $  (\bar u^a_{R}d_{La})$ &$-1$& $-1/2$& $-1/2$\cr
$\Pi^0_d$ & $(\bar d^a_{R}d_{La})$ & $0$ &$-1/2$ &$1/2$\cr
$\Pi^0_u$ & $(\bar u^a_{R}u_{La})$&$ 0$ &$~~1/2$ &$-1/2$\\
\hline
\end{tabular}

\caption{Composite bosons $\Pi^{0,\pm}$, 
their constituents and SM quantum numbers.}\label{qnuP0}
\end{table*}

\begin{figure}   
\begin{center}
\includegraphics[height=4.8cm,width=6.4cm]{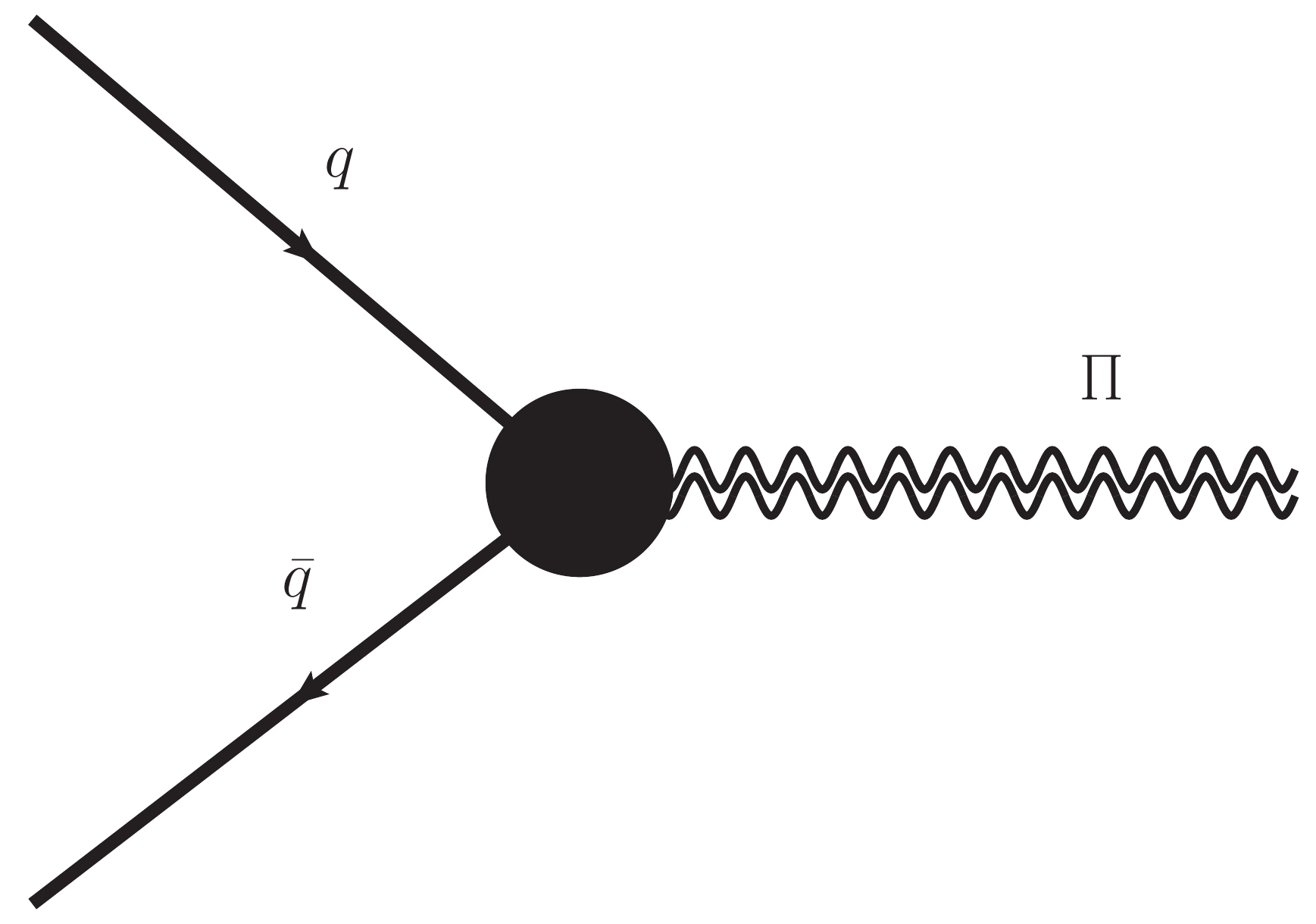}
\vspace{-1em}
\caption{We show the Feynman diagrammatic representation for the contact interaction between the composite boson and its constituent quarks, where the thin solid line represents an SM elementary fermion, double wave line represents a composite boson, and the blob represents an interacting 
vertex $(F_\Pi/\Lambda)^2P_{L,R}$.}
\label{compositeBv}
\end{center}
\vspace{-2em}
\end{figure} 

The contact interactions for the production and decay of a 
composite fermions $F$ are:
\begin{eqnarray}
{\mathcal L}^{F}_{\rm CI} = {\mathcal V}_{F} +{\mathcal V}_{F}^\dagger. 
\label{cieF}
\end{eqnarray}
In the case of Eq.~(\ref{bhlql}) and Table \ref{nuF0},
\begin{eqnarray}
{\mathcal V}_{\bar E^0} &=& \frac{g_*^2}{\Lambda^2}(\bar E^0_{L} e_{R})(\bar d^a_{R}u_{La}), \quad pp~{\rm or}~ep\rightarrow \bar E^0_{L} e_{R},
\label{cieFf0}\\ 
{\mathcal V}_{\bar N^+}&=& \frac{g_*^2}{\Lambda^2}(\bar N^+_{L}\nu^e_{R})(\bar u^a_{R}d_{La}), \quad pp~{\rm or}~ep\rightarrow \bar N^+_{L}\nu^e_{R},
\label{cinuFf0}\\
{\mathcal V}_{\bar E^+} &=& \frac{g_*^2}{\Lambda^2}(\bar E^+_{L}e_{R})(\bar d^a_{R}d_{La}),
\quad pp~{\rm or}~ep\rightarrow \bar E^+_{L}e_{R},
\label{cieF0}\\
{\mathcal V}_{\bar N^0} &=& \frac{g_*^2}{\Lambda^2}(\bar N^0_{L}\nu^e_{R})(\bar u^a_{R}u_{La}), \quad pp~{\rm or}~ep\rightarrow \bar N^0_{L}\nu^e_{R},
\label{cinuF0}
\end{eqnarray}
and their Hermitian conjugates,
\begin{eqnarray}
{\mathcal V}^\dagger_{E^0} &=& \frac{g_*^2}{\Lambda^2}(\bar  e_{L}E^0_{R})(\bar u^a_{R} d_{aL}),\quad E^0_{R}\rightarrow \bar  e_{L}(\bar u^a_{R} d_{aL}),
\label{dcieFf0}\\
{\mathcal V}^\dagger_{N^-}&=& \frac{g_*^2}{\Lambda^2}
(\bar \nu^e_{L}N^-_R)(\bar d^a_{R}u_{aL}),\quad N^-_R\rightarrow\bar \nu^e_{L}(\bar d^a_{R}u_{aL}),
\label{dcinuFf0}\\
{\mathcal V}^\dagger_{E^-} &=& \frac{g_*^2}{\Lambda^2}(\bar e_LE^-_{R})(\bar d^a_{L}d_{Ra}),\quad E^-_{R}\rightarrow \bar e_L(\bar d^a_{L}d_{Ra}),
\label{dcieF0}\\
{\mathcal V}^\dagger_{N^0} &=& \frac{g_*^2}{\Lambda^2}(\bar\nu^e_LN^0_{R})(\bar u^a_{L}u_{Ra}),\quad N^0_{R}\rightarrow \bar\nu^e_L(\bar u^a_{L}u_{Ra}).
\label{dcinuF0}
\end{eqnarray}
In the case of Eq.~(\ref{qbhlql}) and Table \ref{QF0},
\begin{eqnarray}
&&{\mathcal V}_{\bar D^{-2/3}} = \frac{g_*^2}{\Lambda^2}(\bar D^{-2/3}_{Lb} d_{Rb})(\bar d^a_{R}u_{La}); \!\!\!\quad\!\!\! pp~\rightarrow \bar D^{-2/3}_{La} d_{Ra},
\label{cieQf0}\\ 
&&{\mathcal V}_{\bar U^{1/3}}= \frac{g_*^2}{\Lambda^2}(\bar U^{1/3}_{Lb}u_{Rb})(\bar u^a_{R}d_{La}); \quad pp~\rightarrow \bar U^{1/3}_{La}u_{Ra},
\label{cinuQf0}\\
&&{\mathcal V}_{\bar D^{1/3}} = \frac{g_*^2}{\Lambda^2}(\bar D^{1/3}_{Lb}d_{Rb})(\bar d^a_{R}d_{La});
\quad pp~\rightarrow \bar D^{1/3}_{La}d_{Ra},
\label{cieQ0}\\
&&{\mathcal V}_{\bar U^{-2/3}} = \frac{g_*^2}{\Lambda^2}(\bar U^{-2/3}_{Lb}u_{Rb})(\bar u^a_{R}u_{La}); \!\!\!\quad\!\!\!
pp~\rightarrow \bar U^{-2/3}_{Lb}u_{Rb},
\label{cinuQ0}
\end{eqnarray}
and their Hermitian conjugates,
\begin{eqnarray}
\!\!&&{\mathcal V}^\dagger_{D^{2/3}} \!=\! \frac{g_*^2}{\Lambda^2}(\bar  d_{Lb}D^{2/3}_{Rb})(\bar u^a_{R} d_{La}); \!\!\quad\!\!
D^{2/3}_{Rb}\rightarrow \bar  d_{Lb}(\bar u^a_{R} d_{La}),
\label{dcieQf0}\\
\!\!\!&&{\mathcal V}^\dagger_{U^{-1/3}}\! = \!\frac{g_*^2}{\Lambda^2}
(\bar u_{Lb}U^{-1/3}_{Rb})(\bar d^a_{R}u_{La});\!\!\!\quad\!\!\! U^{-1/3}_{Rb}\!\!\rightarrow\!\bar u_{Lb}(\bar d^a_{R}u_{La}),
\label{dcinuQf0}\\
\!\!\!&&{\mathcal V}^\dagger_{D^{-1/3}}\! = \!\frac{g_*^2}{\Lambda^2}(\bar d_{Lb}D^{-1/3}_{Rb})(\bar d^a_{L}d_{Ra});
\!\!\!\quad\!\!\! D^{-1/3}_{Rb}\!\!\rightarrow\! \bar d_{Lb}(\bar d^a_{L}d_{Ra}),
\label{dcieQ0}\\
\!\!&&{\mathcal V}^\dagger_{U^{2/3}} \!=\! \frac{g_*^2}{\Lambda^2}(\bar u_{Lb}U^{2/3}_{Rb})(\bar u^a_{L}u_{Ra});\!\!\quad\!\!
U^{2/3}_{Rb}\rightarrow \bar u_{Lb}(\bar u^a_{L}u_{Ra}).
\label{dcinuQ0}
\end{eqnarray}
These are relevant contact interactions for phenomenological studies of possible inelastic channels of composite-fermion production and decay in  $pp$ or $ep$ collisions.

\begin{figure}   
\begin{center}
\includegraphics[height=4.8cm,width=6.4cm]{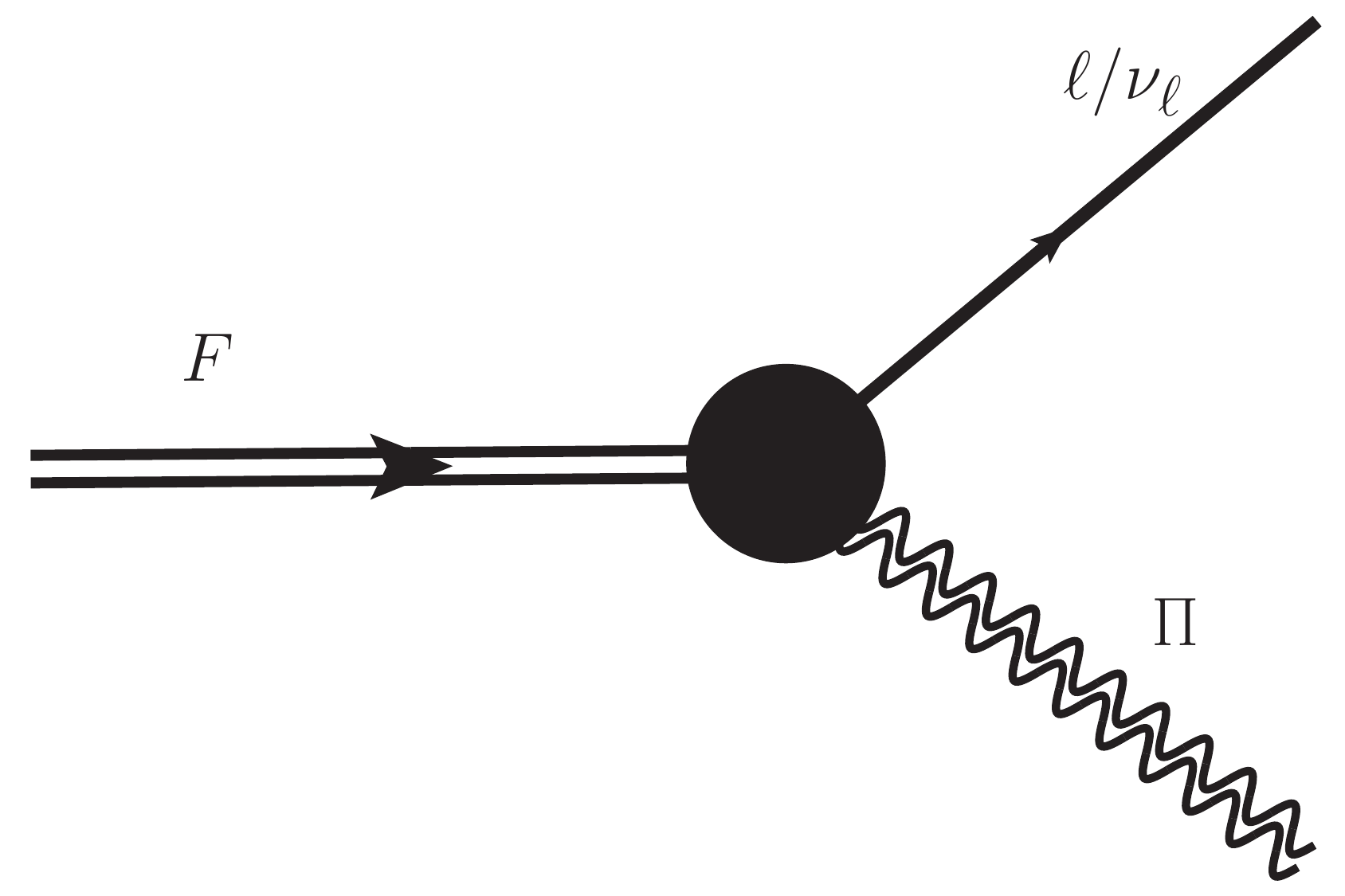}
%
\vspace{-1em}
\caption{We show the Feynman diagrammatic representation for the contact interaction between the composite fermion and boson, 
where the thin solid line represents an SM elementary fermion, the double solid line is a composite fermion and the double wave line represents a composite boson and the blob represents an interacting 
vertex $(F_\Pi/\Lambda)^2P_{L,R}$.}
\label{compositeFBf}
\end{center}
\vspace{-2em}
\end{figure}

\subsection{Composite bosons $\Pi^{0,\pm}$}

From the four-fermion interaction in Eq. (\ref{bhlql}) or (\ref{qbhlql}),  it is possible to form composite bosons
\begin{eqnarray}
&&
\Pi^+=(g^*/F_\Pi)^2(\bar d^a_{R}u_{La}), 
\quad  \Pi^-= (\Pi^+)^\dagger , 
\label{nB0}\\
&&
\Pi^0_d=(g^*/F_\Pi)^2(\bar d^a_{R}d_{La}),\label{dB0}\\
&&
\Pi^0_u=(g^*/F_\Pi)^2(\bar u^a_{R}u_{La}),
\label{uB0}
\end{eqnarray}
and their Hermitian conjugates. 
Such normalized composite boson field has the same dimension 
$[energy]$ of elementary boson field. The composite boson carries the quantum numbers 
that are the sum over SM quantum numbers of its two constituents, 
see Table \ref{qnuP0}. These are pseudo composite bosons $\Pi^{0,\pm}$, analogous to charged and neutral pions $\pi^{0,\pm}$ in the low-energy QCD. 

As shown in Fig.~\ref{compositeBv}, 
the effective coupling between composite boson 
and its two constituents can be written as an effective contact interaction, 
\begin{eqnarray}
{\mathcal L}^{\rm \Pi^\pm}_{\rm CI} &=& g_{_{\rm Y}}(\bar d^a_{R}u_{La})\Pi^- + {\rm h.c.},\label{ciqlB}\\
{\mathcal L}^{\rm \Pi^0_d}_{\rm CI} &=& g_{_{\rm Y}} (\bar d^a_{R} d_{La})\Pi^0_d + {\rm h.c.},
\label{iqlBd}\\
{\mathcal L}^{\rm \Pi^0_u}_{\rm CI} &=& g_{_{\rm Y}}(\bar u^a_{R} u_{Ra})\Pi^0_u + {\rm h.c.},
\label{iqlBu}
\end{eqnarray}
where $g_{_{\rm Y}}=(F_\Pi/\Lambda)^2$.
Appropriately normalizing the composite boson $\Pi$ with the 
form factor $(g^*/F_\Pi)^2$ in Eqs. (\ref{nB0}-\ref{uB0}),  
the effective contact interaction in Eqs. (\ref{ciqlB}-\ref{iqlBu}) can be expressed as a dimensionless Yukawa coupling $g_{_{\rm Y}}$, 
whose value, corresponding to $F_\Pi$ value, 
can be different for composite bosons in Eqs. (\ref{nB0}-\ref{uB0}), but we do not consider such difference here.

\subsection{Contact interaction of composite fermion and boson}
In the view of the composite fermion being a bound state of a composite boson and an SM fermion, using composite-boson fields in Eqs. 
(\ref{nB0}-\ref{uB0}), we rewrite ${\mathcal V}^\dagger$ in Eqs. (\ref{dcieFf0}-\ref{dcinuF0}) as follows, 
\begin{eqnarray}
{\mathcal V}^\dagger_{E^0} &=& g_{_{\rm Y}}(\bar  e_{L}E^0_{R})\Pi^-,\quad E^0_{R}\rightarrow   e_{L}\Pi^+,
\label{pdcieFf0}\\
{\mathcal V}^\dagger_{N^-}&=& g_{_{\rm Y}}
(\bar \nu^e_{L}N^-_R)\Pi^+,\quad N^-_R\rightarrow \nu^e_{L}\Pi^-,
\label{pdcinuFf0}\\
{\mathcal V}^\dagger_{E^-} &=& g_{_{\rm Y}}(\bar e_LE^-_{R})\Pi^0_d,\quad E^-_{R}\rightarrow  e_L\Pi^0_d,
\label{pdcieF0}\\
{\mathcal V}^\dagger_{N^0} &=& g_{_{\rm Y}}(\bar\nu^e_LN^0_{R})\Pi^0_u,\quad N^0_{R}\rightarrow \nu^e_L\Pi^0_u,
\label{pdcinuF0}
\end{eqnarray}
and their Hermitian conjugates ${\mathcal V}$ in Eqs. (\ref{cieFf0}-\ref{cinuF0}), as shown in Fig.~\ref{compositeFBf}. 
In the same way, we rewrite ${\mathcal V}^\dagger$ 
in Eqs. (\ref{dcieQf0}-\ref{dcinuQ0})  as follows, 
\begin{eqnarray}
{\mathcal V}^\dagger_{D^{2/3}} &=& g_{_{\rm Y}}(\bar  d_{L}D^{2/3}_{R})\Pi^-,\quad D^{2/3}_{R}\rightarrow   d_{L}\Pi^+
\label{pdcieQf0}\\
{\mathcal V}^\dagger_{U^{-1/3}}&=& g_{_{\rm Y}}
(\bar u_{L}U^{-1/3}_R)\Pi^+,\quad U^{-1/3}_R\rightarrow u_{L}\Pi^-
\label{pdcinuQf0}\\
{\mathcal V}^\dagger_{D^{-1/3}} &=& g_{_{\rm Y}}(\bar d_LD^{-1/3}_{R})\Pi^0_d,\quad D^{-1/3}_{R}\rightarrow  d_L\Pi^0_d
\label{pdcieQ0}\\
{\mathcal V}^\dagger_{U^{2/3}} &=& g_{_{\rm Y}}(\bar u_LU^{2/3}_{R})\Pi^0_u,\quad U^{2/3}_{R}\rightarrow u_L\Pi^0_u.
\label{pdcinuQ0}
\end{eqnarray}
These contact interactions in Eqs. (\ref{pdcieFf0}-\ref{pdcinuF0}) and (\ref{pdcieQf0}-\ref{pdcinuQ0}) imply that composite fermions $F$: $E^0_R$, $N_R^-$, $E^-_R$, $N_R^0$ and  $F$: $D^{2/3}_R$, $U^{-1/3}_R$, $D^{-1/3}_R$, $U^{2/3}_R$ can decay 
into composite bosons $\Pi^\pm$ and $\Pi^0$, which decay then to 
SM fermions, following the contact interactions in Eqs.~ (\ref{ciqlB}-\ref{iqlBu}) 
at the leading order of tree level. However, we shall consider other decay channels at the next-to-leading order, such as neutral composite boson decay to two SM gauge bosons $\Pi^0_{u,d}\rightarrow \tilde G + \tilde G'$.

\subsection{Contact interaction of $\Pi^0$ composite boson and gauge bosons}\label{spgg}

Analogously to $\pi^0\rightarrow \gamma\gamma$, the massive $\Pi^0_{u,d}$ composite boson can also decay into two gauge bosons \cite{xue2017} :
\begin{eqnarray}
\Pi^0_{u,d} &\rightarrow& \gamma \gamma, \label{cpg}\\
\Pi^0_{u,d} &\rightarrow& \gamma Z^0,\label{cpgz} \\
\Pi^0_{u,d} &\rightarrow& Z^0 Z^0, \label{cpzz} \\
\Pi^0_{u,d} &\rightarrow& W^+ W^-,\label{cpww} 
\end{eqnarray}
via the contact interaction
\begin{eqnarray}
{\mathcal L}^{\Pi^0}_{\tilde G\tilde G'} = \sum_{i=u,d}\frac{gg^\prime N_c}{4\pi^2
F_\Pi}\epsilon_{\mu\nu\rho\sigma}(\partial^\rho A^\mu)(\partial^\sigma A^{\prime\nu})\Pi^0_i,
\label{PiGG} 
\end{eqnarray}
where $g$ and $g^\prime$ represent the couplings 
of gauge bosons $A^\mu$ and $A^{\prime\nu}$ to the SM quarks $u$ 
and $d$ with different $SU_L(2)$-isospin $i=u,d$. 
Actually, this effective contact 
interaction (\ref{PiGG})
is an axial anomaly vertex, as a result of a triangle quark loop and standard renormalization procedure in SM. It should be mentioned that 
two gauge bosons can be two gluons that possibly fuse to a Higgs particle in the final states.

\section{Phenomenology of the composite fermions in $pp$ collisions}\label{PhenomenologyCompositeFermions}

In this section we study the phenomenology of the composite fermions in $pp$ collisions. We first outline their production and decay mode and then calculate their cross section and decay width. This study leads us to the discussion on the search for $F$ that will be discussed in the next section.

\subsection{Production and decay of $F$}
As already specified in Sec.~\ref{contact}, 
the composite fermion $F$ can be 
\begin{eqnarray}
&&E ~(E^0,\bar E^0, E^-,E^+),\nonumber\\
&&N ~(N^0,\bar N^0, N^-,N^+)\nonumber\\
&&D~(D^{2/3},\bar D^{-2/3}, D^{-1/3},\bar D^{+1/3})\nonumber\\ 
&&U~(U^{-1/3},\bar U^{+1/3}, U^{2/3},\bar U^{-2/3}),
\label{compositeSpec}
\end{eqnarray}
where $E,N,D,U$ stand for the charge sector $Q=-1,0,-1/3,2/3$ 
respectively, and the corresponding composite fermions 
of the higher generation of families in the SM.

The kinematics of the processes is derived in the center of mass frame of $pp$ collisions and virtual processes of $F$ are not considered. If the energy $\sqrt{s}$ in the parton center of mass frame is larger than composite fermion masses, the following resonant process can occur:
\begin{eqnarray}
pp &\rightarrow& f ~F
\label{ppfF}
\end{eqnarray}
where the Standard Model fermion $f$ is produced in association with the corresponding composite fermion $F$.
We note that $f$ in Eq.~(\ref{ppfF}) can be $e,\nu,u,d$, and the corresponding Standard Model fermions of the higher generation of families. The kinematics of final states is simple in the center of mass frame of $pp$ collisions.
If we neglect the quark-family mixing, the previous process can manifest at parton level as:
\begin{eqnarray}
u\bar{d} &\rightarrow& e^+ E^0 ~\;{\rm or}\;~ \bar{\nu} N^+ \;~{\rm or}\; ~\bar d D^{2/3},\label{ud}\\
\bar{u}d &\rightarrow& e^- \bar{E^0}~ \;{\rm or}\;~ \nu N^- \;~{\rm or}\;~ \bar u U^{-1/3},\label{udbar}\\
d\bar{d} &\rightarrow& e^{+} E^{-} \;~{\rm or}\;~
e^{-} E^{+}\;~{\rm or}~\bar d D^{-1/3},\label{dd}\\
u\bar{u} &\rightarrow&  \bar{\nu}N^0 \;~ {\rm or}\;~ \nu \bar{N}^0 \;~{\rm or}\;~ \bar u U^{2/3} \label{ddd}.
\end{eqnarray}

The composite fermion $F$ can decay through two different channels: $\bar f$ plus two quarks, via the interactions in Eqs.~(\ref{dcieFf0},~\ref{dcinuFf0},~\ref{dcieF0},~\ref{dcinuF0}) and Eqs.~(\ref{dcieQf0},~\ref{dcinuQf0},~\ref{dcieQ0},~\ref{dcinuQ0}); or $\bar f$ plus a composite boson $\Pi$, 
via the interactions in Eqs.~(\ref{pdcieFf0},\ref{pdcinuFf0},\ref{pdcieF0},\ref{pdcinuF0}) and 
Eqs.~(\ref{pdcieQf0},\ref{pdcinuQf0},\ref{pdcieQ0}, \ref{pdcinuQ0}), 
where $\bar f$ indicates a fermion that is the 
antiparticle of $f$. Then the composite fermion $F$ decays as:
\begin{eqnarray}
F~&\rightarrow& \bar f~ qq^\prime ,\label{Fqq} \\
F~ &\rightarrow& \bar f~ \Pi^{0,\pm}.\label{FP} 
\end{eqnarray}
The full decay chain is:
\begin{eqnarray}
pp &\rightarrow& f F  \rightarrow f \bar f qq^\prime ,\\
pp &\rightarrow& f F \rightarrow f \bar f ~ \Pi^{0,\pm}.
\end{eqnarray}

It appears clear, considering all the possible flavors of $f$ 
and $F$, that a large range of final states is possible. 
The cross section of the process $pp \rightarrow f F $, the decay branching ratio of $F$, and the final states and their topologies are discussed below.

\begin{figure*}[t]
\centering
\includegraphics[scale=1.11]{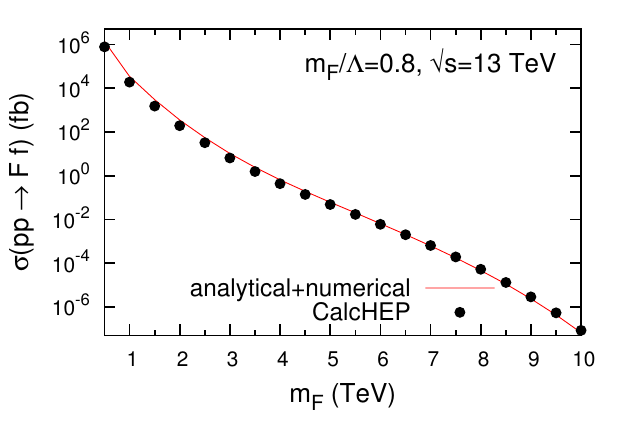}\hspace{0.5cm}
\includegraphics[scale=1.11]{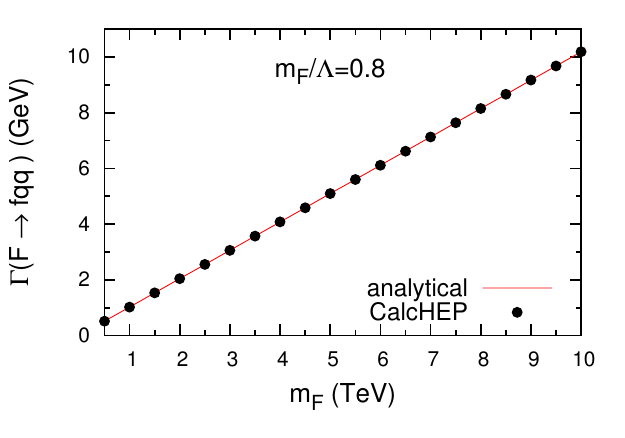}
\caption{\label{ppE0e+} (Color Online). On the left panel, we show the production cross section of 
$pp\rightarrow f F$ as a function of $m_F$ for the case $m_F/\Lambda=0.8$ and at a center of mass energy  $\sqrt{s}=13$ TeV. The solid red line represents the results of an analytical and numerical calculation based on Eqs.~(\ref{csE0}) and (\ref{csFf}) and the filled circles (black) represent the  results from our implementation of the model in CalcHEP. We find good agreement. 
On the right panel, we plot the decay width of composite fermion $F$ as a function of its mass $m_F$ for the case $m_F/\Lambda =0.8$. Again, we observe a good agreement between the expectation from a CalcHEP simulation and the analytical result based on Eq.~(\ref{decayE0}). 
}
\label{E0decaywdth}
\end{figure*} 

\subsection{Cross sections, decay widths and branching ratios}\label{xsec}

\subsubsection{Cross sections}

The partonic cross section of  $qq^\prime\rightarrow f F$ is calculated by standard methods via the contact interaction in Eqs. (\ref{cieFf0}-\ref{dcinuQ0}),
\begin{eqnarray}
\hat\sigma(\hat{s},m_F)= \frac{1}{3\times64\pi}\left(\frac{g_*^2}{\Lambda^2}\right)^2\frac{(\hat{s}-m_F^2)^2}{m_F^2}, 
\label{csE0}
\end{eqnarray}
where $\sqrt{\hat{s}}$ stands for the parton center-mass-energy of $pp$ collisions in LHC experiments.

We consider the production cross section for the composite fermions $F$ in $pp$ collisions expected at the LHC collider according to Feynman's parton model. 
The QCD factorization theorem allows to obtain any hadronic cross section (e.g. in $pp$ collisions) in terms of a convolution of the hard partonic cross sections $\hat\sigma$,
evaluated at the parton center of mass energy $\sqrt{\hat s}=\sqrt{\tau s}$, with the universal parton distribution functions $f_a(x,{\hat{\cal Q}})$ which depend on the parton longitudinal momentum fractions $x$, and on the  factorization scale $\hat{\cal Q}$:
\begin{eqnarray}
\sigma=
\sum_{ij}\!\!\intop^1_{\frac{m_F^{2}}{s}}\!\!d\tau \!\!\intop_\tau^1 \!\!\frac{dx}{x} f_i(x,\hat {\cal Q}^2)f_j(\frac{\tau}{x},\hat {\cal Q}^2)\hat\sigma(\tau s,m_F)\, .
\label{csFf}
\end{eqnarray}
The factorization and renormalization scale ${\cal Q}$ is generally fixed at the value of the mass that is being produced.  
The parametrization of the parton distribution function is NNPDF3.0~\cite{Ball:2014uwa} and the factorization scale has been chosen as $\hat{\cal Q}=m_F$.

The right panel of Fig.~\ref{ppE0e+} shows the agreement between analytical calculations based on Eqs.~(\ref{csE0}) and (\ref{csFf}) 
for the case of the composite fermion $F$, and the results of simulations with CalcHEP where the model with four-fermion interactions has been implemented. 
We remark the quite good agreement that validates our model implementation in CalcHEP.

\subsubsection{Decay widths}\label{xwidth}

Analytical calculations, in the similar way as the first term in Eq. (5) of Ref.~\cite{LPF2014}, yield to the width of the 3-body process $F \rightarrow \bar f qq^\prime$ 
\begin{eqnarray}
\Gamma_{3-\text{body}}(F \to \bar f qq^\prime )= \left(\frac{g_*^2}{\Lambda^2}\right)^2\frac{m_F^5}{4\times(8\pi)^3}. 
\label{decayE0}
\end{eqnarray} 
Note that at TeV energy scales, composite fermions are massive ($m_F$) Dirac fermions, 
whereas all SM elementary fermions are treated as massless Dirac fermions 
of four spinor components, consisting of right- and left-handed Weyl fermions of 
two spinor components. 
Alternatively, the decay width $\Gamma_{F}$ has also been evaluated via CalcHEP, and numerical results are completely in agreement with analytical one in Eq. (\ref{decayE0}), 
see the left panel of Figure \ref{E0decaywdth}.

The decay width of the composite fermion $F$ in the process $F \rightarrow \bar f \Pi$ can easily be computed from the effective contact lagrangian in Eqs.~(\ref{pdcieFf0}-\ref{pdcinuQ0})
\begin{equation}
\Gamma(F \to \bar f \Pi) = \frac{1}{32\pi}\left(\frac{F_\Pi^2}{\Lambda^2}\right)^2 \, m_F\, \left(1-\frac{m_\Pi^2}{m_F^2}\right)^2,
\label{fep}
\end{equation}
and the total width is
\begin{equation}
\Gamma_{\rm tot}(F)=\Gamma(F\rightarrow \bar f\Pi)+\Gamma_{3-body}(F\rightarrow 
\bar fqq^\prime).\label{gtfeqq}
\end{equation}

The decay width of the $\Pi$ boson to two quarks is simply calculated by using 
the effective contact Lagrangian in Eq. (\ref{ciqlB}) and (\ref{iqlBd}),
\begin{eqnarray}
\Gamma(\Pi\to qq^\prime) = \frac{3}{16\pi}\, \left(\frac{F_\Pi}{\Lambda}\right)^4 \,m_\Pi.
\label{Gamma_pqq}
\end{eqnarray} 
For the case that $\Pi$ equals to $\Pi^+$ or $\Pi^-$ composite boson, $\Pi\rightarrow qq^\prime$  of Eq.~(\ref{Gamma_pqq})  
is the only decay channel, see Eqs.~(\ref{nB0}) 
and (\ref{ciqlB}).
The $\Pi^0_{u,d}$ composite bosons, instead, can also decay to two gauge bosons $\tilde G \tilde G'$ (\ref{cpg}-\ref{cpww}), according to the contact interaction (\ref{PiGG}), the corresponding decay widths are \cite{xue2017} :
\begin{eqnarray}
\Gamma_{\Pi^0_{u,d}\rightarrow\gamma\gamma}&=&\left(\frac{5}{9}\right)^2\Gamma, \label{pg}\\
\Gamma_{\Pi^0_{u,d}\rightarrow\gamma Z^0}&=&\frac{1}{\sin^2 2\theta_W}\left(\frac{1}{2}-\frac{5}{9}\sin^2 \theta_W\right)^2\Gamma, \label{pgz}\\
\Gamma_{\Pi^0_{u,d}\rightarrow Z^0 Z^0}&\!=\!&\left(\frac{1/2\!-\!\sin^2\theta_W\!+\!(5/9)\sin^4\theta_W}{\sin^2 2\theta_W}\right)^{\!2}\Gamma, \label{pzz}\\
\Gamma_{\Pi^0_{u,d}\rightarrow W^+ W^-}&=&\left(\frac{1}{8\sin^2\theta_W}\right)^2\Gamma,\label{pww}
\end{eqnarray}
where $\theta_W$ is the Weinberg angle, 
\begin{equation}
\Gamma=\left(\frac{\alpha N_c}{3\pi F_\Pi}\right)^2\frac{m_{\Pi^0_{u,d}}^3}{64\pi}, 
\label{ggrate}
\end{equation}
and the number of colors $N_c=3$.
Total decay rate $\Gamma^{\rm tot}(\Pi^0_{u,d}\rightarrow \tilde G\tilde G')$ is the sum over 
all contributions from  Eqs.~(\ref{pg}-\ref{pww}). 
The total $\Pi^0_{u,d}$-decay rate reads
\begin{equation}
\Gamma_{\rm tot}(\Pi^0_{u,d})=\Gamma(\Pi^0_{u,d}\rightarrow qq') 
+\Gamma^{\rm tot}(\Pi^0_{u,d}\rightarrow \tilde G\tilde G'),
\label{ptotal}
\end{equation}
where $\Gamma(\Pi^0_d\rightarrow qq')$ is given by 
Eq.~(\ref{Gamma_pqq}).
Based on these results, we calculate the exact branching ratios of different channels for different parameters of the model.

\subsubsection{Branching ratios}

The branching ratios of the $\Pi^0_{u,d}$ decay to two quarks
$qq^\prime$, 
\begin{eqnarray}
{\mathcal B}(\Pi^0_{u,d}\rightarrow qq^\prime) = \frac{\Gamma(\Pi^0_{u,d}\rightarrow qq^\prime)}{\Gamma_{\rm tot}(\Pi^0_{u,d})},
\label{pqq}
\end{eqnarray}
and the $\Pi^0_{u,d}$ decay to two gauge bosons $\tilde G\tilde G^\prime$, 
\begin{eqnarray}
{\mathcal B}(\Pi^0_{u,d}\rightarrow \tilde G\tilde G^\prime)=\frac{\Gamma^{\rm tot}(\Pi^0_{u,d}\rightarrow \tilde G\tilde G')}{\Gamma_{\rm tot}(\Pi^0_{u,d})}.
\label{pgg}
\end{eqnarray}
Whereas, the branching ratios of the composite fermion $F$  
decay to $f\Pi$,
\begin{equation}
{\mathcal B}(F\rightarrow f\Pi)=\frac{\Gamma(F\rightarrow f\Pi)}{\Gamma_{\rm tot}(F)}.
\label{bEep}
\end{equation}
The branching ratios of the direct
decay $F\rightarrow \bar fqq'$,
\begin{eqnarray}
{\mathcal B}(F\rightarrow \bar f qq^\prime,{\rm direct})=\frac{\Gamma_{\rm 3-body}(F\rightarrow \bar fqq')}{\Gamma_{\rm tot}(F)},
\label{Eeqqdirect}
\end{eqnarray}
and indirect decay $F\rightarrow \bar f \Pi\rightarrow \bar f qq'$,
\begin{eqnarray}
{\mathcal B}(F\rightarrow \bar f\Pi\rightarrow \bar f qq^\prime)=\frac{
\Gamma(F\rightarrow \bar f \Pi)}{\Gamma_{\rm tot}(F)}~{\mathcal B}(\Pi\rightarrow qq').
\label{Eepqq}
\end{eqnarray}
The sum of these two branching ratios gives 
the total branching ratio ${\mathcal B}(F\rightarrow 
\bar fqq^\prime)$ of
$F$ decaying to $\bar fqq^\prime$,
\begin{eqnarray}
{\mathcal B}(F\rightarrow \bar f
qq^\prime) &=& \Gamma^{-1}_{\rm tot}(F) \times \Big[\Gamma_{3-body}(F\rightarrow \bar fqq^\prime)\nonumber\\
&+&\Gamma(F\rightarrow \bar f\Pi)~{\mathcal B}(\Pi\rightarrow qq^\prime)\Big].
\label{Bfeqq}
\end{eqnarray}
For the case $\Pi^\pm$, ${\mathcal B}(\Pi^\pm\rightarrow qq^\prime)=1$. For the case $\Pi^0_{u,d}$, 
${\mathcal B}(\Pi^0_{u,d}\rightarrow qq^\prime)$ is given 
by Eq.~(\ref{pqq}), and the branching ratios of decay 
$F\rightarrow \bar f \Pi^0_{u,d}\rightarrow \bar f
\tilde G\tilde G^\prime$ is given by
\begin{eqnarray}
{\mathcal B}(F\rightarrow \bar f \Pi^0_{u,d}\rightarrow \bar f
\tilde G\tilde G^\prime) &=& \frac{\Gamma(F\rightarrow \bar f\Pi^0_{u,d})}{\Gamma_{\rm tot}(F)}\nonumber\\
&\times&{\mathcal B}(\Pi^0_{u,d}\rightarrow \tilde G\tilde G^\prime).
\label{Bfegg}
\end{eqnarray}
As a result, the cross sections of these processes are:
\begin{eqnarray}
\sigma(pp\rightarrow  f F \rightarrow \bar f f qq^\prime)&=&\sigma(pp\rightarrow  fF)\nonumber\\
&\times &{\mathcal B}(F\rightarrow \bar fqq^\prime),
\label{xsecfqq}
\end{eqnarray}
and 
\begin{eqnarray}
\sigma(pp\rightarrow  f F \rightarrow \bar f f \tilde G\tilde G^\prime) &=&\sigma(pp\rightarrow  fF)\nonumber\\
&\times& {\mathcal B}(F\rightarrow \bar f \Pi^0_{u,d}\rightarrow \bar f
\tilde G\tilde G^\prime).\nonumber\\
\label{xsecfgg}
\end{eqnarray}
All channels of composite fermion $F$ production and decay 
give the same results at this level of approximation by using contact interactions only.  

For the processes with the $e^+e^-qq^\prime$ final state 
in $pp$ collisions, 
the total cross section is approximately given by
\begin{eqnarray}
\!\!\!\!\!\!\!\sigma(pp\!\rightarrow\! e^+e^-qq^\prime)\approx&&\, \sigma(pp\rightarrow e^+ E^0)\times {\mathcal B}(E^0 \rightarrow e^+ \bar u d)\nonumber\\
+&&\, \sigma(pp\rightarrow e^- \bar E^0)\times {\mathcal B}(\bar E^0 \rightarrow e^- u \bar d)\nonumber \\
+&&\, \sigma(pp\rightarrow e^+ \bar E^-)\times {\mathcal B}(\bar E^- \rightarrow e^- d \bar d)\nonumber \\
+&&\, \sigma(pp\!\rightarrow\! e^-  E^+)\times {\mathcal B}( E^+\! \rightarrow\! e^+ d \bar d),
\label{ppcs}
\end{eqnarray} 
and the total width is
\begin{equation}
\Gamma_{\rm tot}(F)=\Gamma(F\rightarrow e^+\Pi)+\Gamma_{3-body}(F\rightarrow e^+qq^\prime).\label{tfeqq}
\end{equation}
The calculation of these quantities will be given in the next sections.

\begin{figure*}[t]
\includegraphics[scale=0.765]{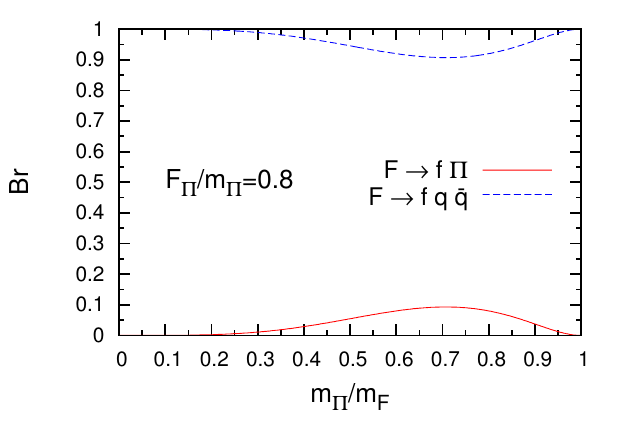}\hspace{0.333cm}
\includegraphics[scale=0.765]{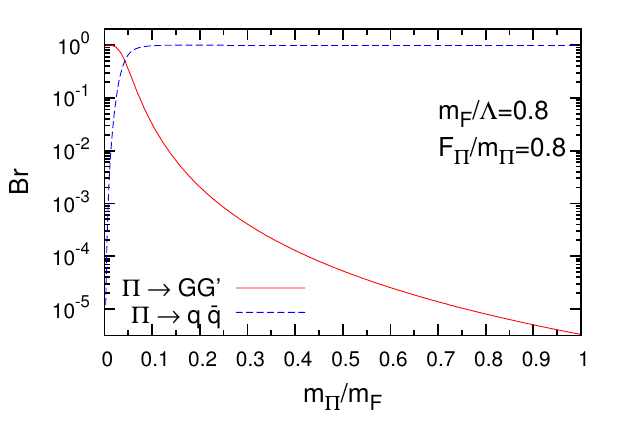}
\includegraphics[scale=0.765]{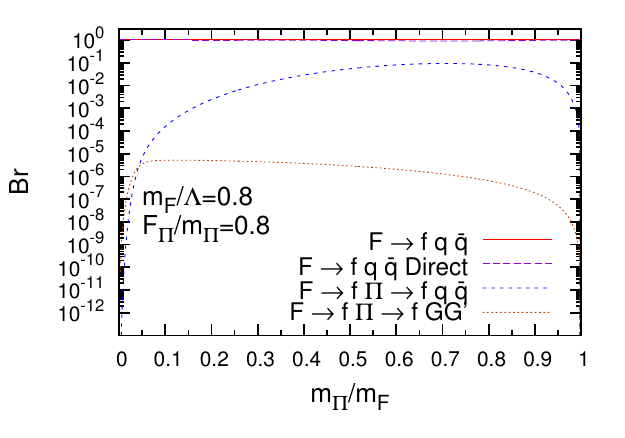}
\includegraphics[scale=0.765]{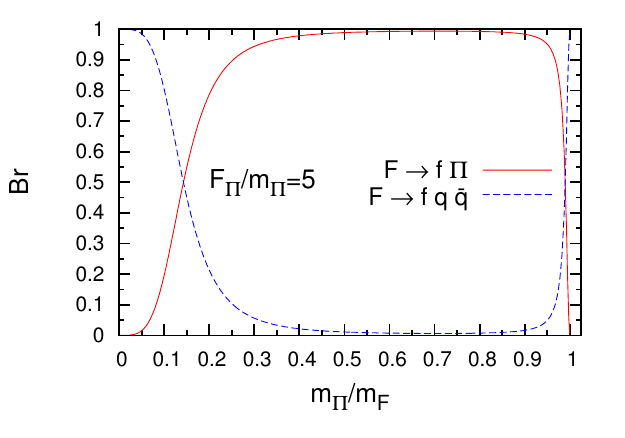}\hspace{0.333cm}
\includegraphics[scale=0.765]{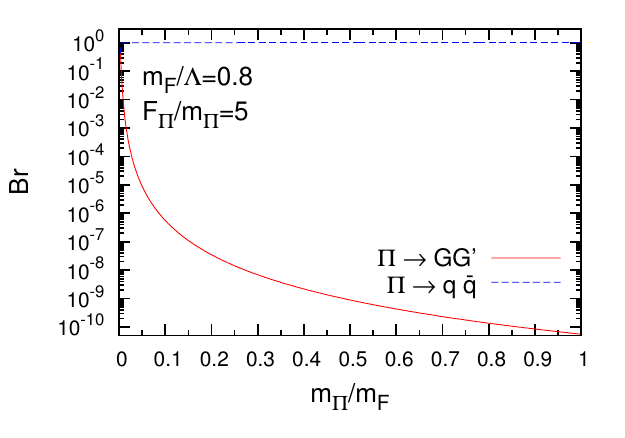}
\includegraphics[scale=0.765]{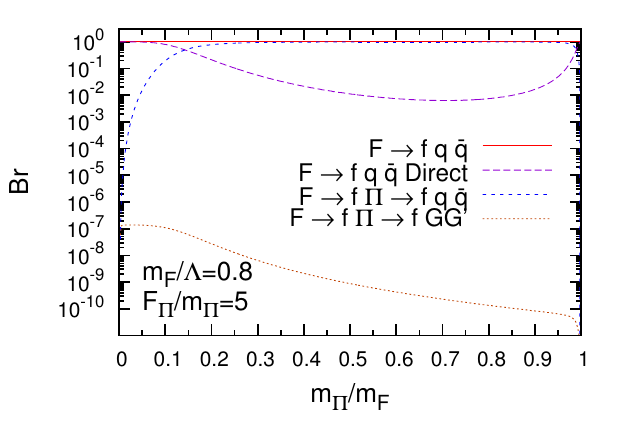}
\includegraphics[scale=0.765]{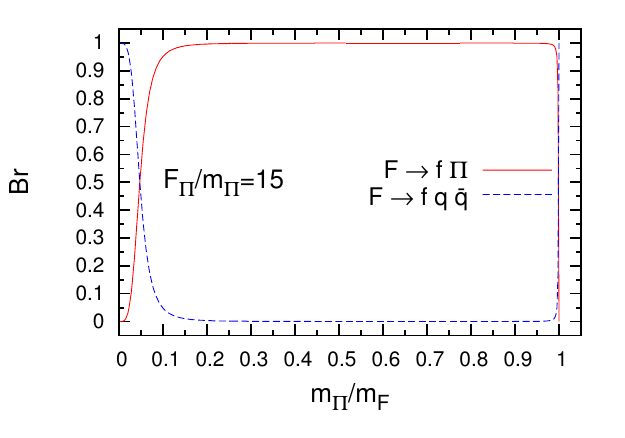}\hspace{0.333cm}
\includegraphics[scale=0.765]{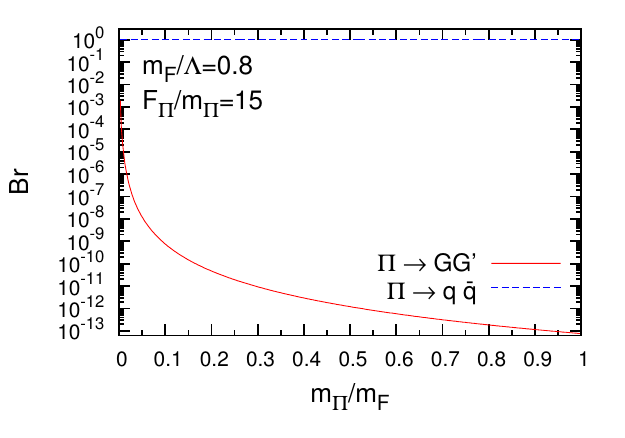}
\includegraphics[scale=0.765]{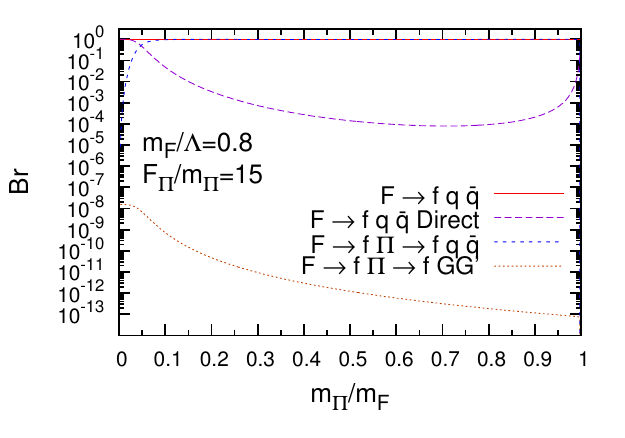}
\caption{\label{AllBr} (Color Online). The branching ratios of $F \rightarrow f qq^\prime$ and $F \rightarrow f \Pi$ (left column), $\Pi \rightarrow qq^\prime$ and $\Pi \rightarrow G\tilde G^\prime$ (center column), and the full decay chains $F \rightarrow f qq^\prime$, $F \rightarrow f \Pi \rightarrow f qq^\prime$, and $F \rightarrow f \Pi \rightarrow f G\tilde G^\prime$ (right column)
are plotted with respect to the ratio $m_\Pi/m_F$ and for the values of 
$m_F/\Lambda=0.8$, $F_\Pi/\,m_\Pi=0.8$ (the 1st row), $F_\Pi/\,m_\Pi=5$ (the 2nd row), $F_\Pi/\,m_\Pi=15$ (the 3rd row). Further plots in a wider range of parameter space
are in Appendix \ref{Br_Fpimpi_1_15_step1}, Fig. \ref{BrAppendix}.
}
\end{figure*}

\comment{ 
\begin{figure*}[t]   
\centering
\caption{\label{AllBr0.2} (Color Online). The branching ratios of $\Pi \rightarrow qq^\prime$ and $\Pi \rightarrow G\tilde G^\prime$ (left column), and the full decay chains $F \rightarrow \bar f qq^\prime$, $F \rightarrow \bar f \Pi \rightarrow \bar f qq^\prime$, and $F \rightarrow \bar f \Pi \rightarrow \bar f G\tilde G^\prime$ (right column)
are plotted with respect to the ratio $m_\Pi/m_F$ and for the values of $m_F/\Lambda=0.2$, $F_\Pi/\,m_\Pi=100$ (the 1st row), $F_\Pi/\,m_\Pi=10$ (the 2nd row), $F_\Pi/\,m_\Pi=1$ (the 3rd row), and $F_\Pi/\,m_\Pi=0.1$ (the 4th row). \red{Note that 
in figures the notation $E$ ($e$) generally stands for a composite fermion $F$ (an SM fermion $f$).}}
\end{figure*}
}

\section{Search for $F$ at the LHC}
\label{SearchForFAtTheLHC}
After having discussed the production and decay of $F$ and its cross-section, width, and decay branching ratio, we now examine these results in terms of parameters of the model and derive the possible final states and their topologies, highlighting their impact to the current program of beyond SM searches at the LHC.

\subsection{Branching ratios and topology of $F$ with respect to model parameters}\label{bran}
In order to present the branching ratios of different possible 
channels in terms of parameters of the model, 
we are bound to discuss physically sensible 
parameters to explore. This model has four parameters that can be rearranged to three dimensionless parameters for a given $\Lambda$ value:
$
(\Lambda, m_F, F_\Pi, m_\Pi) \rightarrow 
( m_F/\Lambda, F_\Pi/m_\Pi, m_\Pi/m_F).
$

The ratio $m_F/\Lambda < 1$ ($m_\Pi/\Lambda < 1$)
of the composite fermion (boson) mass and the basic composite scale 
$\Lambda$ gives us an insight into the dynamics of composite fermion (boson) 
formation. 
In addition, as the parameters $m_\Pi$  and $F_\Pi$ represent the same dynamics of composite boson formation we use the $F_\Pi/m_\Pi$ ratio.
Finally, to take into account the feature that a composite fermion $F$ is composed by a composite boson and an elementary SM fermion, we adopt the ratio  $m_\Pi/m_F < 1$ as a parameter.  
As a result, for given $\sqrt{s}$ and $\Lambda$ values, we have three parameters 
$m_F/\Lambda$, $F_\Pi/ m_\Pi$, and $m_\Pi/m_F$ to represent the results of 
the possible branching ratios. 
Figure \ref{AllBr} shows three sets of plots for the branching ratios of $F \rightarrow \bar f qq^\prime$ and $F \rightarrow \bar f \Pi$ (left column), $\Pi \rightarrow qq^\prime$ and $\Pi \rightarrow G\tilde G^\prime$ (center column), and the full decay chains 
$F \rightarrow \bar f qq^\prime$, $F \rightarrow \bar f \Pi \rightarrow \bar f qq^\prime$, and $F \rightarrow \bar f \Pi \rightarrow \bar f G\tilde G^\prime$ (right column). These branching ratios are plotted with respect to the $m_\Pi/m_F$ ratio and for the values of $F_\Pi/\,m_\Pi=0.8$, $F_\Pi/\,m_\Pi=5$, $F_\Pi/\,m_\Pi=15$. 
Note that 
the branching ratios of $F \rightarrow f qq^\prime$ and $F \rightarrow f \Pi$ (left column) do not depend on $m_F/\Lambda$, which can be seen from Eqs.~(\ref{decayE0},\ref{fep}) and (\ref{gtfeqq}). For the $\Pi^\pm$ decay, the channel (\ref{Gamma_pqq}) is unique, so the branching ratio is one, independent of $m_F/\Lambda$. Whereas the $\Pi^0_{u,d}$ decays also to $G^\prime G$, see Eqs.~(\ref{pqq}) and (\ref{pgg}), and the branching ratio depends on $m_F/\Lambda$. However, in the regime of $m_\Pi/m_F$ we consider, the branching ratio of $\Pi^0_{u,d}\rightarrow G^\prime G$ is very small and negligible, compared with the branching ratio of $\Pi^0_{u,d}\rightarrow q^\prime q$. Therefore, our results of branching ratio of $F$ decay presented in Fig.~\ref{AllBr} 
are independent of the parameter $m_F/\Lambda$. As a result, regarding the branching ratio and topologies of $F$ decay, the model effectively depends only on two parameters $F_\Pi/m_\Pi$ and $m_\Pi/m_F$. However, the cross section of $F$ production depends on the parameter $m_F/\Lambda$, as indicated in Eq. \ref{csE0}.

In Fig.~\ref{AllBr}, 
it is shown that the branching ratios $F\rightarrow f q\bar q ~{\rm direct}$ (\ref{Eeqqdirect}) and $F\rightarrow f\Pi \rightarrow f q\bar q~{\rm indirect}$ (\ref{Eepqq}) 
tend to swap each other for different values of $F_\Pi/m_\Pi$. 
Increasing the coupling $F_\Pi$ (and the ratio $F_\Pi/m_\Pi$) of composite boson $\Pi$ to its two constituents  ($q\bar q$), see 
contact interactions (\ref{ciqlB},\ref{iqlBd},\ref{iqlBu}), 
the branching ratios $F\rightarrow f\Pi \rightarrow f q\bar q~{\rm indirect}$ (\ref{Eepqq}) 
becomes dominant over 
the branching ratios $F\rightarrow f q\bar q ~{\rm direct}$ (\ref{Eeqqdirect}). We thus consider two reference cases of $F_\Pi/m_\Pi$: 
\begin{enumerate}[(i)]
\item $F_\Pi/m_\Pi\lesssim 0.8$ where $F\rightarrow f q\bar q ~{\rm direct}$ dominates. We have verified that for the value 0.8 the direct production dominates by at least a factor 10 over the production with the $\Pi$ for all values of $m_\Pi/m_F$.
\item $F_\Pi/m_\Pi\gtrsim 5$ where $F\rightarrow f\Pi \rightarrow f q\bar q~{\rm indirect}$ (\ref{Eepqq}) becomes relevant for all values of $m_\Pi/m_F$ and dominates above $m_\Pi/m_F>$ 0.2 over $F\rightarrow f q\bar q ~{\rm direct}$. We notice from Fig.~\ref{AllBr} 
that as $F_\Pi/m_\Pi$ increases above 15, the decay $F\rightarrow f\Pi \rightarrow f q\bar q~{\rm indirect}$ dominates over all possible decays for all values of $m_\Pi/m_F$. 
\end{enumerate}

The expected topologies related to the phenomenology of $F$ are summarized in Table \ref{FinalStatesTopologiesF0}, considering the two complementary cases of $F_\Pi/m_\Pi\leq 0.8$ and $F_\Pi/m_\Pi\geq 15$. These cases are representative of the topology of $F$ in all the phase space, including the intermediate region $0.8 < F_\Pi/m_\Pi < 15$, in which the decay $F\rightarrow f q\bar q ~{\rm direct}$ or $F\rightarrow f\Pi \rightarrow f q\bar q~{\rm indirect}$ dominates depending on the specific value of $F_\Pi/m_\Pi$ considered. To this purpose we provide in Appendix~\ref{Br_Fpimpi_1_15_step1} the total decay branching ratios corresponding to Fig.~\ref{AllBr} (right column) for values of $F_\Pi/m_\Pi$ between 1 and 14, increasing in step of 1. We further outline that the value 0.2 for $m_\Pi/m_F$ that separates the boosted from the resolved topology when $F$ decays through a $\Pi$ is indicative and may vary based on the mass of $F$. We have verified, using CalcHEP, that for a mass of $F$ of 1 TeV, the two quarks decaying from the $\Pi$ originating from $F$ are indeed within a $\Delta R(q,\bar{q}')$ below 0.8, which determines the cone size of a large-radius jets suitable for boosted jet at the LHC experiments. For a mass of $F$ of 7 TeV instead, the value 0.2 is lowered to 0.15 to guarantee $\Delta R(q,\bar{q}')<0.8$, as for higher masses of $F$ its width increases.

\begin{table*}[!]
\addtolength{\tabcolsep}{2pt}

\scalebox{0.95}{\begin{tabular}{ccccccc}
\hline
$F_\Pi/m_\Pi$ & $m_\Pi/m_F$ &  Channel & Resonances & Topology & Experimental features\\
\hline
15  & $[\sim 0.2, \sim 1]$ & $f F \rightarrow f (\bar f \Pi) \rightarrow f (\bar f (qq^\prime))$ & $F, \Pi$ & Resolved w/ $\Pi \rightarrow qq^\prime$ &  identification of $\Pi$ and $F$\\
    & $\leq 0.2$ & $f F \rightarrow f (\bar f \Pi) \rightarrow f (\bar f (qq^\prime))$ & $F$, $\Pi$ & Boosted & identification of $F$;   & \\
    & & & & & $\Pi$ large-radius jet: \\
    & & & & & 2-prong, no V boson tag  \\
\hline
$\leq$ 0.8 & [0,1] & $f F \rightarrow f (\bar f qq^\prime)$ & $F$ & Fully resolved &  same of $F_\Pi/m_\Pi$ = 10\\
\hline
\end{tabular}}

\vskip-5pt
\caption{Summary of the relevant channels where $F$ can decay considering two complementary values of $F_\Pi/ m_\Pi$ and the ranges of $m_\Pi/m_F$ that correspond to different topologies of $F$. Similar considerations on the channels where $F$ can decay and its topology apply to all values of $F_\Pi/ m_\Pi$, depending on the value of $F_\Pi/ m_\Pi$, the value of $m_\Pi/m_F$, and the branching ratios of $F\rightarrow f q\bar q ~{\rm direct}$ and $F\rightarrow f\Pi \rightarrow f q\bar q~{\rm indirect}$.}
\label{FinalStatesTopologiesF0}
\end{table*}

\begin{table*}[!]
\addtolength{\tabcolsep}{5pt}

\scalebox{0.925}{\begin{tabular}{ccccccc}
\hline
\hline
 $f$    & $F$ & Topology &  Final state & LHC searches & Features not exploited in LHC searches\\
\hline 
 $e$    & $E$ 
  & Fully resolved & $e^\pm (e^\mp qq^\prime)$ & \cite{LARFP2017,llqq_FR_ATLAS} & $E$ identification & \\ 
& & Resolved w/ $\Pi \rightarrow qq^\prime$  & $e^\pm (e^\mp (qq^\prime))$ & \cite{llqq_Rpi_CMS,llqq_FR_ATLAS} & $E,\Pi$ identification & \\ 
& & Boosted  & $e^\pm e^\mp J$ & \cite{LARFP2017} & 2-prong, no V boson tag,  boosted $\Pi$  decay & \\
 \hline 
 $\mu$  & $M$ 
  & Fully resolved & $\mu^\pm (\mu^\mp qq^\prime)$ & \cite{LARFP2017,llqq_FR_ATLAS} & $M$ identification & \\ 
& & Resolved w/ $\Pi \rightarrow qq^\prime$ & $\mu^\pm (\mu^\mp (qq^\prime))$ &   \cite{llqq_Rpi_CMS,llqq_FR_ATLAS} & $M, \Pi$ identification & \\ 
& & Boosted  & $\mu^\pm \mu^\mp J$ & \cite{LARFP2017} & 2-prong, no V boson tag,  boosted $\Pi$  decay & \\
 \hline 
 $\tau$ & $\mathcal{T}$ 
  & Fully resolved & $\tau^\pm (\tau^\mp qq^\prime)$ & \cite{tautauqq_FR_CMS} & $\mathcal{T}$ identification & \\ 
& & Resolved w/ $\Pi \rightarrow qq^\prime$ & $\tau^\pm (\tau^\mp (qq^\prime))$ & \cite{tautauqq_FR_CMS} & $\mathcal{T},\Pi$ identification & \\ 
& & Boosted  & $\tau^\pm \tau^\mp J$ & n/a & & \\
 \hline 
 $\nu$  & $N$ 
  & Fully resolved & $\nu (\nu qq^\prime)$ & \cite{nunuqq_FR_CMS,nunuqq_FR_ATLAS} & $N$ identification & \\ 
& & Resolved w/ $\Pi \rightarrow qq^\prime$ & $\nu (\nu (qq^\prime))$ & \cite{nunuqq_FR_CMS,nunuqq_FR_ATLAS} & $N,\Pi$ identification & \\ 
& & Boosted  & $\nu \nu J$ & \cite{nunuJ_boost_ATLAS} & 2-prong, no V boson tag,  boosted $\Pi$ deacy & \\
 \hline 
 $j$    & $J$ 
  & Fully resolved & $j (j qq^\prime)$ & n/a & & \\
& & Resolved w/ $\Pi \rightarrow qq^\prime$ & $j (j (qq^\prime))$ & n/a & & \\
& & Boosted  & $j j J$ & n/a & & \\
 \hline 
 $c$    & $C$ 
  & Fully resolved & $c (c qq^\prime)$ & n/a & & \\
& & Resolved w/ $\Pi \rightarrow qq^\prime$ & $c (c( qq^\prime))$ & n/a & & \\
& & Boosted  & $c c J$ & n/a & & \\
 \hline 
 $b$    & $B$ 
  & Fully resolved & $b (b qq^\prime)$ & n/a & & \\
& & Resolved w/ $\Pi \rightarrow qq^\prime$ & $b (b( qq^\prime))$ & n/a & & \\
& & Boosted  & $b b J$ & n/a & & \\
 \hline 
 $t$    & $T$ 
  & Fully resolved & \textbf{$t (\bar{t} qq^\prime)$} & n/a & & \\
& & Resolved w/ $\Pi \rightarrow qq^\prime$ & \textbf{$t (\bar{t} (qq^\prime))$} & n/a & & \\
& & Boosted  & $t \bar{t} J$ & n/a & & \\
\hline
\end{tabular}}

\vskip-5pt
\caption{List of all the signatures foreseen by the model for the different flavors of $f$, $F$, their topology and final states. In the table ``$j$''  indicates any reconstructed jet that originates from a $u,d,s$ quark and ``$J$'' a large-radius (cone-size of 0.8) jet that is reconstructed from 2 quarks produced with low angular separation. The notation w$/\Pi$ indicates the decay of $F$ through a $\Pi$. The LHC searches for these signatures are reported along with some comments on the features that are typical of this model and have not been exploited in the referenced LHC searches. In the table, "n/a" indicates the case in which an LHC search interesting the corresponding final state has not been found.}
\label{FinalStatesTopologiesF}
\end{table*}

So far we have presented the studies relying on the first SM generation, i.e. considering the electron flavoflavorr of $F$ ($F = E$). However, these discussions and calculations are straightforwardly generalized to the second and third SM generations, and in the text we keep the general notation $F$ instead of $E$. At the 
leading order of only contact interactions being taken into account, the formulae of cross-sections, decay rates and 
branching ratios of the second and third generations are the same as those of the first generation, however composite fermions $F$ and their productions, decay channels and rates are different, depending 
on the values of their parameters $m_F$, $m_\Pi$ and $F_\Pi$, which
vary from one SM generation to another.
The reasons are that the effective couplings $\sim g_*^2/\Lambda^2$ and 
$\sim g_{_Y}$ of contact interactions can be different, due to the 
modifications from the flavor mixing and gauge interactions. 
However, we neglect these modifications 
in this article, and expect the variations of masses 
$m_F$, $m_\Pi$, decay 
constant $F_\Pi$ to be small. In this sense, there are two basic 
parameters $F_\Pi/m_\Pi$ and $m_\Pi/m_F$ 
for each SM generation to be determined by different channels and 
their topologies in LHC experiments discussed in next sections.

\subsection{Signatures to search for $F$ at the LHC}


We now summarize on the possible signatures with which $F$ can manifest at the LHC. We remark that $F$ can have the different flavors corresponding to the Standard Model flavors of $f$ and the fact that these particles are not necessarily mass degenerate. This implies that they can have different masses with which they can appear within the energy reach of the LHC, and thus they have to be searched for independently. Based on the flavors of $f$, we expect 8 different final states to be investigated for the pair $fF$ produced in the process $pp\rightarrow fF$, which are: $eE,\mu M,\tau\mathcal{T},\nu N,jJ,cC,bB,tT$. Here, we consider one single channel ($\nu N$) for all the $\nu$ neutrinos of the Standard Model and one single channel ($jJ$) for the $u,d,s$ quarks. We notice that the flavor $cC$ is taken separately, because of the improving performances in c-tagging algorithms at the LHC (\cite{cTagATLAS, cTagCMS}) and dedicated searches for new physics with $c$ quarks in the final state (\cite{cTagSearchATLAS, cTagSearchCMS}). Moreover, we distinguish the three topologies (resolved with and without $\Pi$ and boosted) explained in the previous paragraph, so that we have in total 24 different signatures that have to be considered in order to pursue a comprehensive search of $F$. 

In Table ~\ref{FinalStatesTopologiesF} we outline these signatures based on the flavors $f$ and $F$, the possible topology, the corresponding final states, and the LHC search that, to the best of our knowledge, could be more sensitive to searching for $F$. In the last column, we further report on features of $F$ and its decay that have not been exploited directly in the cited LHC searches and could be used to improve possible future searches dedicated to $F$. We especially remark that the $F$ quark flavors appear to be completely unexplored yet and we urge on the importance of carrying out specific analyses at LHC to investigate it. This is certainly noteworthy and can have a relevant impact on the beyond the SM physics program of the LHC.

We notice that, despite possible and with a peculiar signature, the channel $f F \rightarrow f (\bar f \Pi) \rightarrow f (\bar f (\tilde G\tilde G^\prime))$ is negligible since the decay $\Pi^0 \rightarrow (\tilde G\tilde G^\prime)$ is only relevant for $F_\Pi/\,m_\Pi=0.1$ when, however, the decay $F \rightarrow f (\Pi)$ has a branching ratio close to zero. Because of this we do not include this case in Table ~\ref{FinalStatesTopologiesF} and we point out that this case would become of interest in the case $F$ is found in one of the possible signatures mentioned above, to study the nature of the new particle. \comment{It should be mentioned that these gauge bosons $\tilde G\tilde G'$ in Eq.~(\ref{PiGG}) can be two gluons that possibly fuse to a Higgs particle in the final states. Do we keep this part? how is it related to BSM searches? should we mention about future studies on this case?} 

We put emphasis on the case of the composite fermions 
$F=N^0,\bar N^0, N^+,N^-$ for the final state $\nu\nu qq^\prime$, where $\nu\bar\nu$ stands for the pairs of the SM left-handed neutrino $\nu^e_L$ 
and/or sterile right-handed neutrino $\nu^e_R$, as the latter is a candidate of dark-matter particles. 

Finally, we acknowledge that the final state $f \bar f qq^\prime$ is relevant for a wide range of the parameter space of the model and thus will consider it in the next section and, in particular, the case of $f = e$ and $F = E$ to derive limits on the model parameters based on existing LHC results.

\section{Bounds on the model parameters}
\label{bound}
\begin{figure}[ht]
\centering
\includegraphics[scale=1.35]{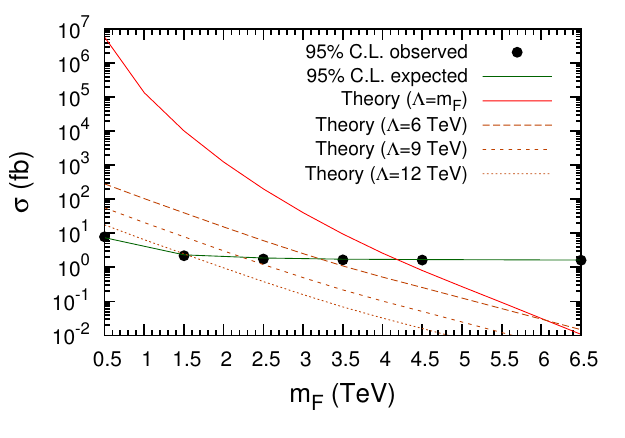}
\caption{\label{recast} (Color Online). Recast of the experimental upper limit on $\sigma (pp \to eeqq^\prime)$ published in \cite{LARFP2017} against the model of composite fermions studied in this article. The dotted line (solid green line) 
is the 95\% C.L. observed (expected) upper limit on $\sigma (pp \to eeqq^\prime)$ as reported in \cite{LARFP2017}. The solid line (red) is the theoretical expectation from the model described in this work as given by Eq.~(\ref{ppcs})  
for the case $m_F/\Lambda=1$, the dashed lines (orange) are the theoretical expectation from the model for the cases $\Lambda=6,\,9,\,12$ TeV. If $m_F/\Lambda=1$ one obtains that the composite fermions of this model are excluded up to masses $m^{\rm ex}_F\approx 4.25$ TeV.}
\label{1dimlimit}
\end{figure} 

\begin{figure}[ht] 
\centering
\includegraphics[scale=1.35]{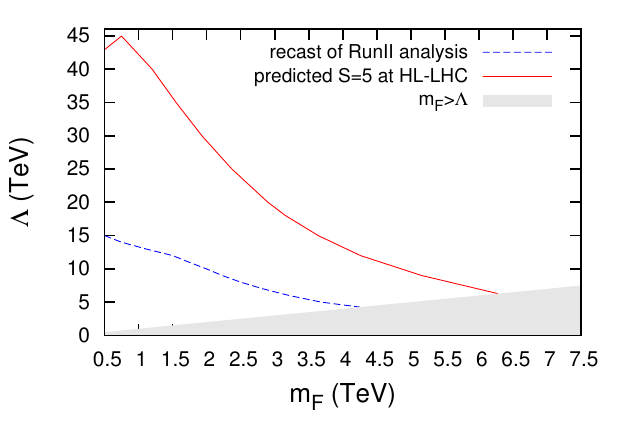}
\caption{\label{recast2dim} (Color Online). Recast of the experimental upper limit from \cite{LARFP2017} (dashed line) and the predicted contour curve at a 5-level statistical significance (solid line) in the 2-dimensional parameter space 
$(\Lambda,m_F)$. 
The  shaded region denotes unphysical values of the parameters ($\Lambda < m_F$). 
}
\label{2dimlimit}
\end{figure}

In this section we provide a discussion of the bounds of the model parameters taking as a reference the case $F = E$ and the final state $eeqq^\prime$ that has been shown to be sensitive to a wide portion of the parameter space of the model in the previous section. For this purpose, we
recast the 95\% confidence level (C.L.) experimental upper limit on $\sigma(pp\rightarrow eeqq^\prime)$ using a recent analysis~\cite{LARFP2017} 
of 2.3 fb$^{-1}$ data from the 2015 Run II of the LHC by the CMS collaboration with respect to the predictions of the  model of composite fermions discussed in this article. Note that both electrons and positrons are collected in the final states of $eeqq^\prime$, electrons and positrons are not distinguished  in the data analysis.
For the case $m_F/\Lambda=1$ one obtains that the composite 
fermions of this model are excluded up to masses 
$m^{\rm ex }_F\approx 4.25$ TeV. 
This result is shown in Figure \ref{recast}, together with the exclusion limits 
$m^{\rm ex }_F\approx 3.3, 2.4,1.5$ TeV for $\Lambda$ fixed at 6, 9 and 12 TeV. 
Figure \ref{2dimlimit} shows the exclusion curve, lower (dashed) line, in the 2-dimensional parameter space $(\Lambda,m_F)$ for the model obtained via  the recasting of the analysis~\cite{LARFP2017} 
of 2.3 fb$^{-1}$ data from the 2015 Run II of the LHC by the CMS collaboration. Here the regions of the parameter space below the curves are excluded. 

We also performed a study about the potential of a dedicated analysis in the High Luminosity LHC (HL-LHC) conditions (center of mass energy of 14 TeV and luminosity of 3 ab$^{-1}$). We used CalcHEP to generate the processes and DELPHES \cite{Delphes} to simulate the detector effects. In order to separate the signal from the background, we selected events with $pt_{e_1}\geq$ 180 GeV, $pt_{e_2}\geq$ 80 GeV, $pt_{j_1}\geq$ 210 GeV, $m_{ee}\geq$ 300 GeV ($pt$ is the transverse momentum, $e_1$ the leading electron, $e_2$ the subleading electron, $j_1$ the leading jet and $m_{ee}$ the invariant mass of the two electrons). Then we evaluated the reconstruction and selection efficiencies for signal ($\epsilon_s$) and background ($\epsilon_b$) as the ratio of the selected and the total generated events. From these efficiencies, the signal and  background cross sections ($\sigma_s$, $\sigma_b$) and the integrated luminosity ($L$), it is possible to evaluate the expected number of events for the signal ($N_s$) and the SM background ($N_b$) and finally the statistical significance ($S$):
\begin{equation}
N_s=L\sigma_s\epsilon_s, \quad N_b=L\sigma_b\epsilon_b, \quad S=\frac{N_s}{\sqrt{N_b}}\, .
\end{equation}
The $S=5$ contour curve is shown by the upper (solid) line in Figure \ref{2dimlimit}. It can be used to get indications about the potential for discovery or exclusion with the experiments at the HL-LHC, showing that there is a wide region of the model phase space where the existence of the composite fermions can be investigated;
for the case $m_F/\Lambda=1$ we can reach masses up to $\approx 6.2$ TeV.
We notice that, despite having considered the case of $F = E$ in this section, it could be inferred that the cross section for $F$, which should approximately be the same for all its flavors, is sufficient for the $F$ to appear at the LHC with the statistics already collected at the LHC experiments, and that is expected by the HL-LHC. Based on this result, we recommend that the physics program of the LHC consider the new particles foreseen by this model and their signatures in its investigations.

\section{Summary and remarks}
\label{SummaryAndRemarks}

In the  weak coupling regime the effective four-fermion operators of NJL-type 
possess an IR-fixed point, rendering the elegant Higgs mechanism of the SM of particle physics at low energies. In the strong coupling regime, on the other end, these operators could possess an UV-fixed point, giving rise to composite 
fermions $F$ (bosons $\Pi$) composed by SM fermions as bound states of three (two) SM elementary leptons or quarks, and to their relevant contact interactions with them at high energies ${\mathcal O}({\rm TeV})$.

We study, for the first time for this model, the spectrum of composite particles and contact interactions in quark-lepton and quark-quark sectors in relationship to their phenomenology at the LHC in order to unveil their discovery potential. The cross sections and decay rates of composite particles are calculated based on the LHC physics from $pp$ collision at high energy TeV scale. We find out that a comprehensive investigation of the model presented here can be effectively achieved, for given $\sqrt{s}$  and $\Lambda$ values, by considering only two parameters: $F_\Pi/\,m_\Pi$ and $m_\Pi/m_F$. Based on these results, we exhaustively examine all the possible $F$ states and the signatures with which they can manifest at the LHC, according to different $F_\Pi/\,m_\Pi$ and $m_\Pi/m_F$. Interestingly, we find that there is a broad variety of new composite particles that could manifest in signatures that have escaped the realm of the searches at the LHC. We summarize these cases in Table ~\ref{FinalStatesTopologiesF}. They can offer an unprecedented discovery potential of physics beyond the SM and we urge on the importance for the LHC experiments to include such searches in their ongoing physics program.

In order to set bounds on the model parameters, we derive constraints for the particular case where $F$ has electron-like flavor. We analyzed the particular processes giving $e^+e^- qq'$ final state by using the recast of the experimental upper limit by the CMS collaboration on the cross-section $\sigma(pp\rightarrow eeqq')$. We determine that a composite fermion $F$ of mass $m_F$ below 4.25 TeV can be excluded for $\Lambda$ = $m_F$. At the same time, we compute 3$\sigma$ and 5$\sigma$ contour plots of the statistical significance and highlight the phase space in which $F$ can manifest using 3 ab$^{-1}$, foreseen at the high luminosity LHC (HL-LHC). This result shows that, even for final state traditionally considered at the LHC experiments, there is a vast range of model parameters to which a dedicated search can be sensitive to the $F$ composite fermions. We encourage such efforts in future investigations in light of peculiar features of $F$ not yet exploited at the LHC searches and highlighted in Table ~\ref{FinalStatesTopologiesF}.
We are preparing the next article presenting the 
investigation of phenomenology at the LHC of composite bosons.

Both composite bosons ($\Pi$) and fermions ($F$) are mass 
eigenstates and have definite SM quantum numbers, so that the CKM 
flavor mixing and the Feynman diagrammatic representations of SM 
perturbative gauge interactions can be easily implemented, see 
Eqs.~(4.8)-(4.11) in Ref.~\cite{xue2017}. At the leading order 
of contact interactions considered in this article, the contributions from 
all CKM flavor mixing and gauge interactions to composite particle masses 
are neglected. It should be mentioned that these effective couplings between composite particles and SM gauge bosons have two main effects. First, they give the possible 
decay channels of composite particles to final states involving SM gauge bosons, like 
two gauge bosons as in Sec.~\ref{spgg}. The branching ratios of these 
channels are negligible in the composite fermion production and decay 
studied in this article. However, they could be important in the composite boson formation and decay, which is under examination in a separate effort.  
Second, these effective couplings between composite particles and SM gauge bosons 
should affect the well-measured SM quantities in the IR regime, like the electroweak oblique parameters or the decay width of the Z boson, as well as the Higgs physics. These issues are indeed important and necessary, and will be addressed in a future study.
So far only some general and qualitative discussions have been presented \cite{xue2016_2,xue2017}, showing that the composite Higgs boson is a 
tight bound state of $t \bar t$, as if it was an elementary particle,  
and possible corrections to the SM quantities are small. 

It is an interesting question to see how these phenomenologies can 
possibly account for some recent results obtained in 
both space and underground laboratories. The cosmic rays $pp$ collisions might
produce composite fermions $F=E$ that decay into electrons 
and positrons. This may explain an excess of cosmic ray electrons and 
positrons around TeV scale \cite{changjin2008,dark2009}. 
In addition, recent AMS-02 results \cite{AMS2015} 
show that at TeV scale the energy-dependent proton flux changes 
its power-law index. This implies that there would be ``excess'' 
TeV protons whose origin could be also explained by the resonance of  
composite fermions $F=N$ due to the interactions of dark-matter and 
normal-matter particles. These composite fermions should appear as 
resonances by high-energy sterile neutrinos inelastic collisions 
with nucleons (xenon) at the largest cross-section, then resonances decay 
and produce some other detectable SM particles in underground 
laboratories \cite{pandaX}. Similarly, in the ICECUBE 
experiment \cite{icecube}, we expect events where the neutrinos change 
their directions (lower their energies) by their inelastic collisions to form the
resonances of composite fermions $N$ at a high energy scale ($\approx$ TeV). It is worthwhile to mention that 
in the IR domain of this model there are effective coupling 
vertexes of the SM gauge boson $W$ and the right-handed currents: 
$g_R\bar\nu_R\gamma^\mu e_R W^+_\mu$ or $g_R\bar u_R\gamma^\mu d_R W^+_\mu$, where $g_R\propto (v/\Lambda)^2$ \cite{xueRcoupling}. 
The parity symmetry is restored at the scale $\Lambda$ 
\cite{xue2016_2,xue2017} and none of additional intermediate 
gauge bosons $W_R$ or $W'$ is present. 
The recent phenomenological studies of this effective coupling 
in the quark sector can be found in Ref.~\cite{Rcurrent2017}.
In the lepton sector, these effective contact interactions 
relate to the 
dark-matter physics of right-handed neutrinos $\nu_R$. 
Similarly to the analogy between the 
Higgs mechanism and BCS superconductivity, the composite-particle 
counterparts in condensed matter physics have been recently 
discussed \cite{KX2017}.

\begin{acknowledgments}

The work of Alfredo Gurrola and Francesco Romeo is supported in part by NSF Award PHY-1806612. The work of Hao Sun is supported by the National Natural Science Foundation of China (Grant No.11675033).
\end{acknowledgments}
\appendix
\section{$F$ branching ratios for $F_\Pi/m_\Pi$ assuming values in [1,15]}\label{Br_Fpimpi_1_15_step1}
In this appendix we provide the branching ratios of all the possible decays of $F$, as in  Fig.~\ref{AllBr} (right column), for values of $F_\Pi/m_\Pi$ between 1 and 15, increasing in step of 1.

\begin{figure*}[h] 
\centering
\includegraphics[scale=0.775]{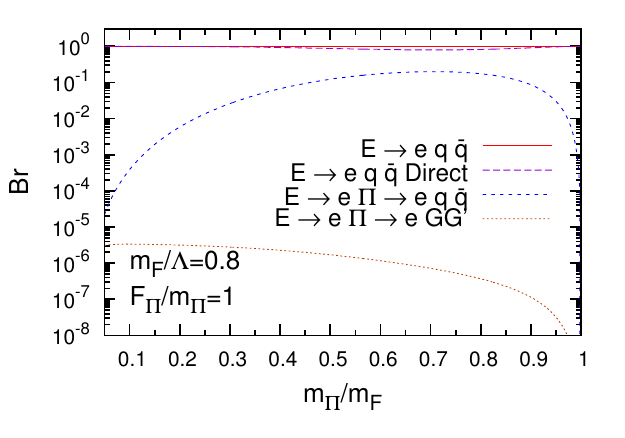}
\includegraphics[scale=0.775]{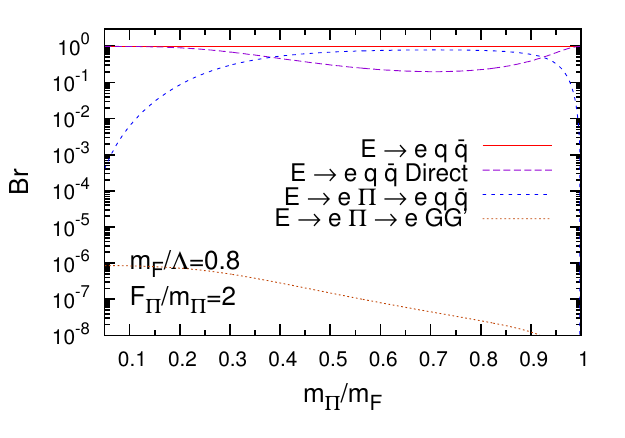}
\includegraphics[scale=0.775]{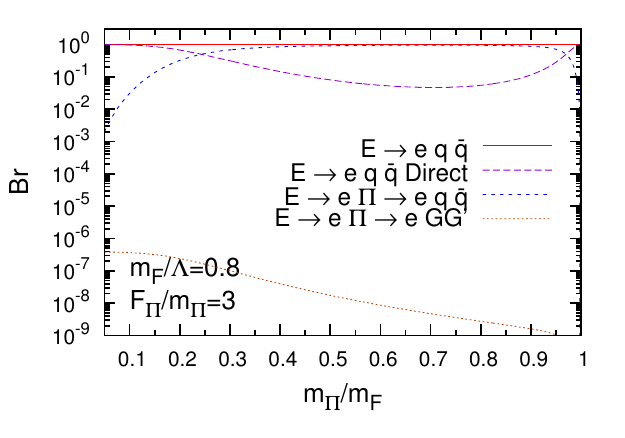}
\includegraphics[scale=0.775]{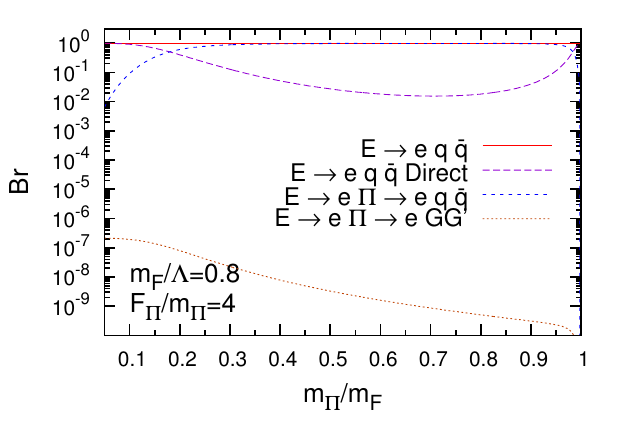}
\includegraphics[scale=0.775]{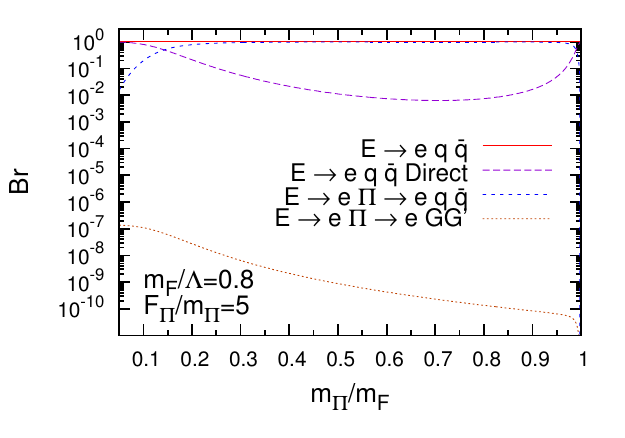}
\includegraphics[scale=0.775]{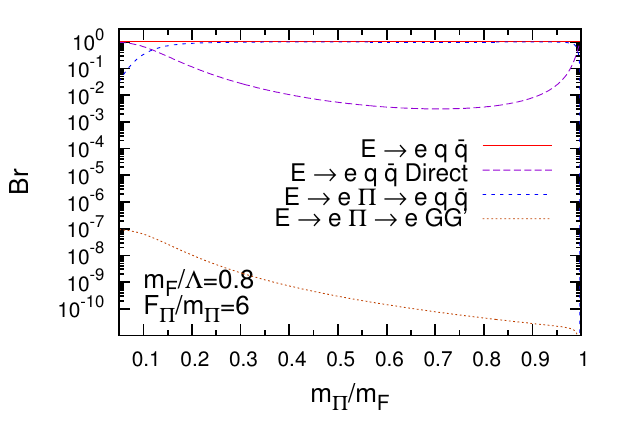}
\includegraphics[scale=0.775]{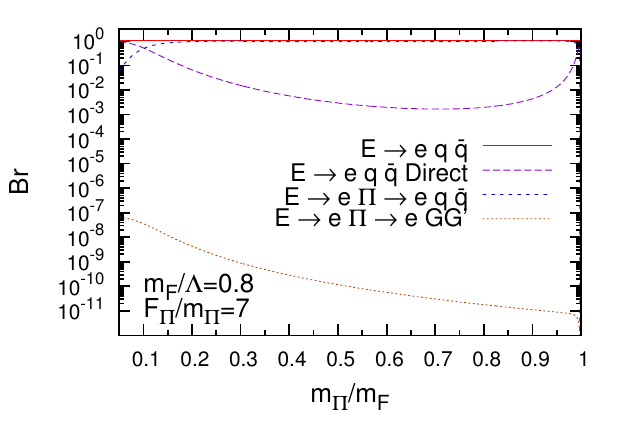}
\includegraphics[scale=0.775]{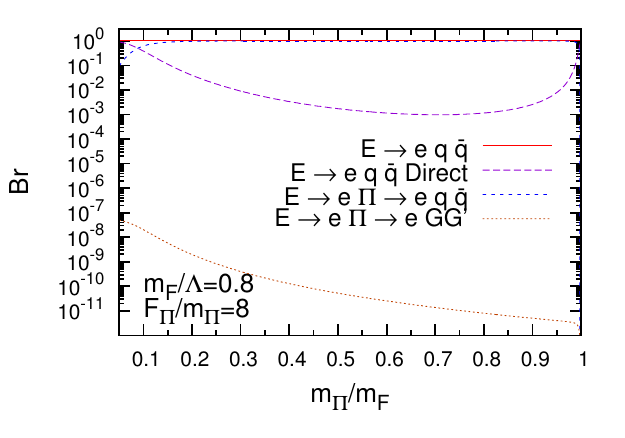}
\includegraphics[scale=0.775]{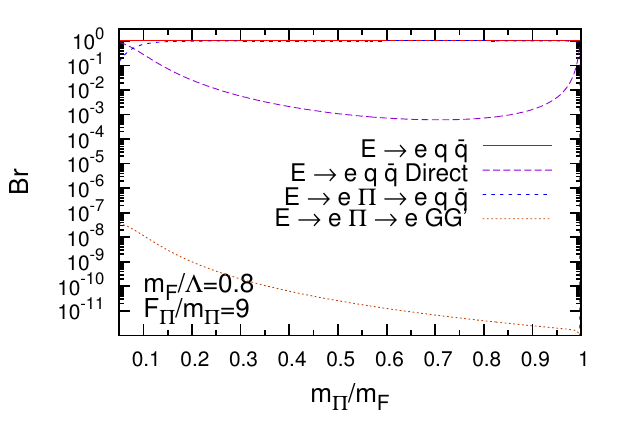}
\includegraphics[scale=0.775]{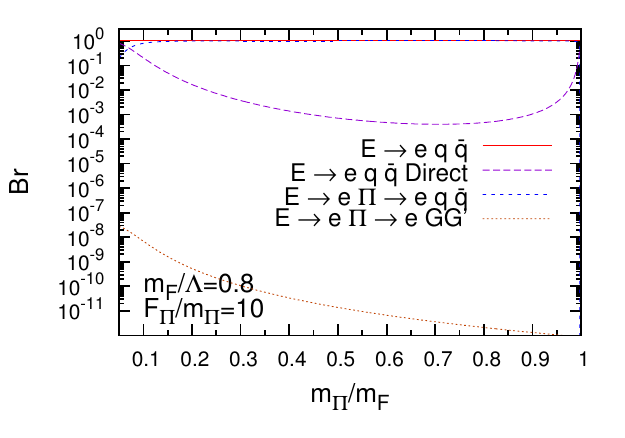}
\includegraphics[scale=0.775]{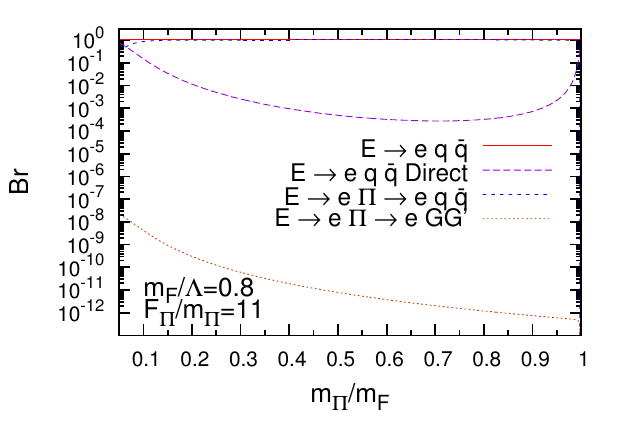}
\includegraphics[scale=0.775]{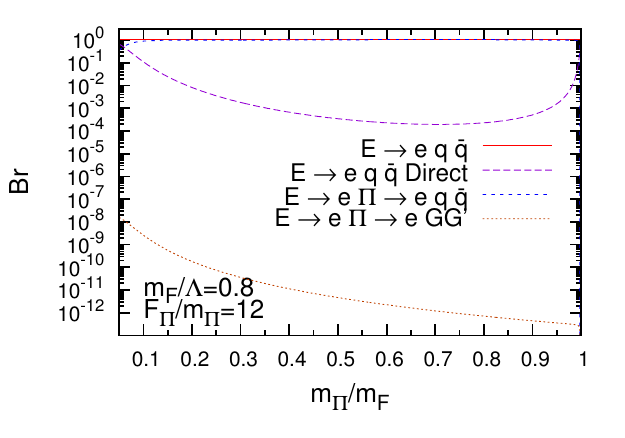}
\includegraphics[scale=0.775]{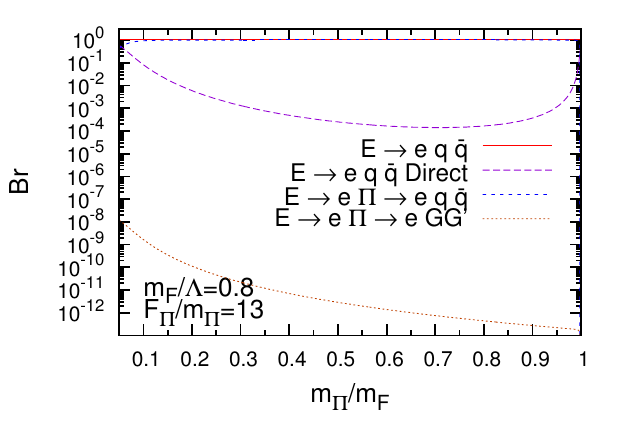}
\includegraphics[scale=0.775]{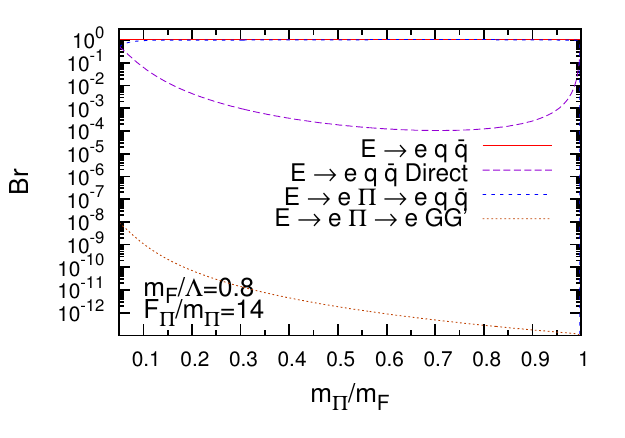}
\includegraphics[scale=0.775]{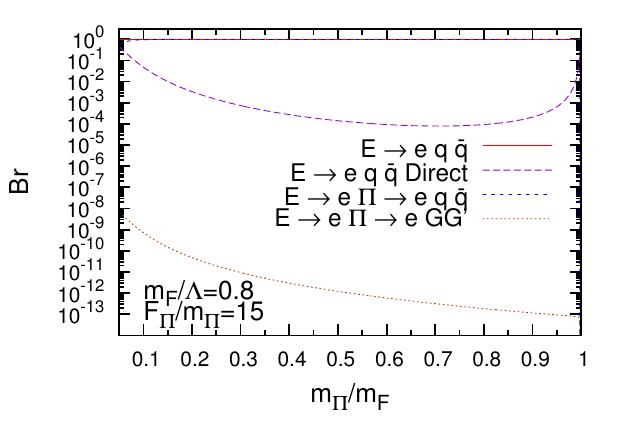}
\caption{\label{BrAppendix} (Color Online). The branching ratios of all the possible decays of $F$ for values of $F_\Pi/m_\Pi$ between 1 and 15, increasing in step of 1, to complement the cases reported in Fig.~\ref{AllBr} (right column).}
\end{figure*}

\end{document}